\documentclass[pra, aps, preprint, superscriptaddress, showkeys, a4paper, hidelinks,nofootinbib]{revtex4-2} %reprint, preprint
\usepackage[T1]{fontenc}
\usepackage{aas_macros}
\usepackage{lmodern}
\usepackage[official,right]{eurosym}
\usepackage{textcomp}   % provides proper micro- and degree-symbol
\usepackage{gensymb}
\usepackage[english]{babel}
\usepackage[pdftex]{graphicx,epsfig}
\usepackage{rotating}
\usepackage{eso-pic}
\usepackage{dsfont}\let\mathbb\mathds
\usepackage[breaklinks=true,hidelinks]{hyperref}
\usepackage[section]{placeins} %allows to have unlimited floats
\usepackage{needspace}
\usepackage{setspace}
\usepackage[suffix='']{epstopdf}
\usepackage{tabularx}
\usepackage{subfig}
\usepackage{multirow}
\usepackage{MnSymbol}
\usepackage{xspace}
\usepackage{siunitx}%added by Martin
\usepackage[normalem]{ulem}%addeded by PJC for strikethrough text
\usepackage{footmisc}%for \footref
\graphicspath{{./Images/}}

\newlength{\bildtitel}
\newcommand\REVIEW[1]{\message{LaTeX Warning: \noexpand untreated nEDM-REVIEW command in \jobname .tex: l\the\inputlineno}}% for publication
\newcommand{\diff}[1]{\operatorname{d}\ifthenelse{\equal{#1}{}}{\,}{\!#1}}

\newcommand{\ecm}{\ensuremath{\si{\elementarycharge}\!\cdot\!\cm}}

\newcommand{\Nup}{\ensuremath{N^{\uparrow}}}
\newcommand{\Ndo}{\ensuremath{N^{\downarrow}}}

\newcommand{\cm}{\ensuremath{\mathrm{cm}}}
\newcommand{\mm}{\ensuremath{\mathrm{mm}}}

\newcommand{\lampHg}{\texorpdfstring{\ensuremath{{}^{204}\text{Hg}}}{mercury-204}\xspace}
\newcommand{\magHg}{\texorpdfstring{\ensuremath{{}^{199}\text{Hg}}}{mercury-199}\xspace}

\newcommand{\Rr}[1]{\ifthenelse{\equal{#1}{}}{\ensuremath{\mathcal{R}}}{\ensuremath{\mathcal{R}_{#1}}}}

\newcommand{\Tp}{\ensuremath{\mathcal{T}}} % free precession time cf. normal T for temperature 

\newcommand{\Gz}{\ensuremath{G_{\rm grav}}}

\newcommand{\Bo}{\ensuremath{\vec{B}_0}}

\newcommand{\fN}{\ensuremath{f_{\rm n}}}
\newcommand{\fHg}{\ensuremath{f_{\rm Hg}}}

\newcommand{\bUcn}{\ensuremath{b_{\rm UCN}^{\ast}}} 
\newcommand{\bAlps}{\ensuremath{b_{\rm ALP}^{\ast}}} 
\newcommand{\mAlps}{\ensuremath{m_{\rm ALP}}} 

\newcommand{\gN}{\ensuremath{\gamma_{\rm n}}}
\newcommand{\gHg}{\ensuremath{\gamma_{\rm Hg}}}

\newcommand{\mN}{\ensuremath{m_{\rm n}}}

\newcommand{\Nbot}{\ensuremath{N_{\rm bot}}}
\newcommand{\Ntop}{\ensuremath{N_{\rm top}}}

\newcommand{\rhoUcn}[1]{\ifthenelse{\equal{#1}{}}{\ensuremath{\rho_{\rm n}}}{\ensuremath{\rho_{\rm n}\left(#1\right)}}}

\begin{document}

\title{Search for an interaction mediated by axion-like particles with ultracold neutrons at the PSI}

%\author{\color{red}C.~Abel}
%\affiliation{Department of Physics and Astronomy, University of Sussex, Falmer, Brighton BN1 9QH, United Kingdom}

\author{N.~J.~Ayres}
\affiliation{ETH Z\"{u}rich, Institute for Particle Physics and Astrophysics, CH-8093 Z\"{u}rich, Switzerland}

%\author{\color{red}G.~Ban}	
%\affiliation{Normandie {Universit\'e}, ENSICAEN, UNICAEN, CNRS/IN2P3, LPC Caen, 14000 Caen, France}

\author{G.~Bison}
\affiliation{Paul Scherrer Institut, CH-5232 Villigen PSI, Switzerland}

\author{K.~Bodek}
\affiliation{Marian Smoluchowski Institute of Physics, Jagiellonian University, 30-348 Cracow, Poland}

\author{V.~Bondar}
\affiliation{ETH Z\"{u}rich, Institute for Particle Physics and Astrophysics, CH-8093 Z\"{u}rich, Switzerland}

\author{T.~Bouillaud}
\affiliation{Universit\'e Grenoble Alpes, CNRS, Grenoble INP, LPSC-IN2P3, 38026 Grenoble, France}

\author{E.~Chanel}
\altaffiliation[Present address: ]{Institut Laue Langevin, 38000 Grenoble, France}
\affiliation{Laboratory for High Energy Physics and Albert Einstein Center for Fundamental Physics, University of Bern, CH-3012 Bern, Switzerland}

\author{P.-J.~Chiu}
\email[Corresponding author: ]{pin-jung.chiu@physik.uzh.ch}
\altaffiliation[Present address: ]{University of Zurich, Switzerland}
\affiliation{ETH Z\"{u}rich, Institute for Particle Physics and Astrophysics, CH-8093 Z\"{u}rich, Switzerland}
\affiliation{Paul Scherrer Institut, CH-5232 Villigen PSI, Switzerland}	

\author{B.~Clement}
\affiliation{Universit\'e Grenoble Alpes, CNRS, Grenoble INP, LPSC-IN2P3, 38026 Grenoble, France}

\author{C.~B.~Crawford}
\affiliation{Department of Physics and Astronomy, University of Kentucky, Lexington, Kentucky, 40506, USA}

\author{M.~Daum}
\affiliation{Paul Scherrer Institut, CH-5232 Villigen PSI, Switzerland}

%\author{\color{red}B.~Dechenaux}	
%\affiliation{Normandie {Universit\'e}, ENSICAEN, UNICAEN, CNRS/IN2P3, LPC Caen, 14000 Caen, France}

\author{C.~B.~Doorenbos}
\affiliation{ETH Z\"{u}rich, Institute for Particle Physics and Astrophysics, CH-8093 Z\"{u}rich, Switzerland}
\affiliation{Paul Scherrer Institut, CH-5232 Villigen PSI, Switzerland}	

\author{S.~Emmenegger}
\affiliation{ETH Z\"{u}rich, Institute for Particle Physics and Astrophysics, CH-8093 Z\"{u}rich, Switzerland}

%\author{\color{red}L.~Ferraris-Bouchez}
%\affiliation{Universit\'e Grenoble Alpes, CNRS, Grenoble INP, LPSC-IN2P3, 38026 Grenoble, France}

\author{M.~Fertl}
\affiliation{Institute of Physics, Johannes Gutenberg University Mainz, 55128 Mainz, Germany}

\author{P.~Flaux}	
\affiliation{Normandie {Universit\'e}, ENSICAEN, UNICAEN, CNRS/IN2P3, LPC Caen, 14000 Caen, France}

%\author{\color{red}A.~Fratangelo}
%\affiliation{Laboratory for High Energy Physics and Albert Einstein Center for Fundamental Physics, University of Bern, CH-3012 Bern, Switzerland}

\author{W.~C.~Griffith}
\affiliation{Department of Physics and Astronomy, University of Sussex, Falmer, Brighton BN1 9QH, United Kingdom}

%\author{\color{red}Z.~D.~Gruji\'{c}}
%\affiliation{Physics Department, Universit\'{e} de Fribourg, CH-1700 Fribourg, Switzerland}
%\affiliation{Institute of Physics Belgrade, University of Belgrade, 11080 Belgrade, Serbia}

\author{P.~G.~Harris}
\affiliation{Department of Physics and Astronomy, University of Sussex, Falmer, Brighton BN1 9QH, United Kingdom}

\author{N.~Hild}
\affiliation{ETH Z\"{u}rich, Institute for Particle Physics and Astrophysics, CH-8093 Z\"{u}rich, Switzerland}
\affiliation{Paul Scherrer Institut, CH-5232 Villigen PSI, Switzerland}	

\author{M.~Kasprzak}
\affiliation{Paul Scherrer Institut, CH-5232 Villigen PSI, Switzerland}	

\author{K.~Kirch}
\affiliation{ETH Z\"{u}rich, Institute for Particle Physics and Astrophysics, CH-8093 Z\"{u}rich, Switzerland}
\affiliation{Paul Scherrer Institut, CH-5232 Villigen PSI, Switzerland}	

\author{V.~Kletzl}
\affiliation{ETH Z\"{u}rich, Institute for Particle Physics and Astrophysics, CH-8093 Z\"{u}rich, Switzerland}
\affiliation{Paul Scherrer Institut, CH-5232 Villigen PSI, Switzerland}	

%\author{\color{red}P.~Knowles}
%\affiliation{Physics Department, Universit\'{e} de Fribourg, CH-1700 Fribourg, Switzerland}

\author{P.~A.~Koss}
%\altaffiliation[Present address: ]{Fraunhofer-Institut f\"{u}r Physikalische Messtechnik IPM, 79110 Freiburg i. Breisgau, Germany}
\affiliation{Instituut voor Kern- en Stralingsfysica, University of Leuven, B-3001 Leuven, Belgium}

\author{J.~Krempel}
\affiliation{ETH Z\"{u}rich, Institute for Particle Physics and Astrophysics, CH-8093 Z\"{u}rich, Switzerland}

\author{B.~Lauss}
\affiliation{Paul Scherrer Institut, CH-5232 Villigen PSI, Switzerland}

\author{T.~Lefort}	
\affiliation{Normandie Universit\'e, ENSICAEN, UNICAEN, CNRS/IN2P3, LPC Caen, 14000 Caen, France}

%\author{\color{red}Y.~Lemiere}	
%\affiliation{Normandie Universit\'e, ENSICAEN, UNICAEN, CNRS/IN2P3, LPC Caen, 14000 Caen, France}

%\author{\color{red}A.~Leredde}
%\affiliation{Universit\'e Grenoble Alpes, CNRS, Grenoble INP, LPSC-IN2P3, 38026 Grenoble, France}

\author{P.~Mohanmurthy}
%\altaffiliation[Present address: ]{University of Chicago, 5801 S Ellis Ave, Chicago, IL 60637, USA.}
\altaffiliation[Present address: ]{Massachusetts Institute of Technology, MA, USA}
\affiliation{ETH Z\"{u}rich, Institute for Particle Physics and Astrophysics, CH-8093 Z\"{u}rich, Switzerland}
\affiliation{Paul Scherrer Institut, CH-5232 Villigen PSI, Switzerland}	

\author{O.~Naviliat-Cuncic}
\affiliation{Normandie Universit\'e, ENSICAEN, UNICAEN, CNRS/IN2P3, LPC Caen, 14000 Caen, France}

\author{D.~Pais}
\affiliation{ETH Z\"{u}rich, Institute for Particle Physics and Astrophysics, CH-8093 Z\"{u}rich, Switzerland}
\affiliation{Paul Scherrer Institut, CH-5232 Villigen PSI, Switzerland}	

\author{F.~M.~Piegsa}
\affiliation{Laboratory for High Energy Physics and Albert Einstein Center for Fundamental Physics, University of Bern, CH-3012 Bern, Switzerland}

\author{G.~Pignol}
\affiliation{Universit\'e Grenoble Alpes, CNRS, Grenoble INP, LPSC-IN2P3, 38026 Grenoble, France}

\author{M.~Rawlik}
\affiliation{ETH Z\"{u}rich, Institute for Particle Physics and Astrophysics, CH-8093 Z\"{u}rich, Switzerland}

%\author{\color{red}D.~Rebreyend}
%\affiliation{Universit\'e Grenoble Alpes, CNRS, Grenoble INP, LPSC-IN2P3, 38026 Grenoble, France}

\author{I.~Rien\"{a}cker}
\affiliation{ETH Z\"{u}rich, Institute for Particle Physics and Astrophysics, CH-8093 Z\"{u}rich, Switzerland}
\affiliation{Paul Scherrer Institut, CH-5232 Villigen PSI, Switzerland}	

\author{D.~Ries}
\affiliation{Department of Chemistry - TRIGA site, Johannes Gutenberg University Mainz, 55128 Mainz, Germany}

\author{S.~Roccia}
\affiliation{Universit\'e Grenoble Alpes, CNRS, Grenoble INP, LPSC-IN2P3, 38026 Grenoble, France}

\author{D.~Rozpedzik}
\affiliation{Marian Smoluchowski Institute of Physics, Jagiellonian University, 30-348 Cracow, Poland}

\author{P.~Schmidt-Wellenburg}
\email[Corresponding author: ]{philipp.schmidt-wellenburg@psi.ch}
\affiliation{Paul Scherrer Institut, CH-5232 Villigen PSI, Switzerland}

\author{N.~Severijns}
\affiliation{Instituut voor Kern- en Stralingsfysica, University of Leuven, B-3001 Leuven, Belgium}

\author{B.~Shen}
\affiliation{Department of Chemistry - TRIGA site, Johannes Gutenberg University Mainz, 55128 Mainz, Germany}

\author{K.~Svirina}
\affiliation{Universit\'e Grenoble Alpes, CNRS, Grenoble INP, LPSC-IN2P3, 38026 Grenoble, France}

\author{R.~Tavakoli~Dinani}
\affiliation{Instituut voor Kern- en Stralingsfysica, University of Leuven, B-3001 Leuven, Belgium}

\author{J.~A.~Thorne}
\affiliation{Laboratory for High Energy Physics and Albert Einstein Center for Fundamental Physics, University of Bern, CH-3012 Bern, Switzerland}

\author{S.~Touati}
\affiliation{Universit\'e Grenoble Alpes, CNRS, Grenoble INP, LPSC-IN2P3, 38026 Grenoble, France}

%\author{R.~Virot}
%\affiliation{Universit\'e Grenoble Alpes, CNRS, Grenoble INP, LPSC-IN2P3, 38026 Grenoble, France}

\author{A.~Weis}
\affiliation{Physics Department, Universit\'{e} de Fribourg, CH-1700 Fribourg, Switzerland}

\author{E.~Wursten}
\altaffiliation[Present address: ]{RIKEN, Wako, Saitama 351-0198, Japan}
\affiliation{Instituut voor Kern- en Stralingsfysica, University of Leuven, B-3001 Leuven, Belgium}

\author{N.~Yazdandoost}
\affiliation{Department of Chemistry - TRIGA site, Johannes Gutenberg University Mainz, 55128 Mainz, Germany}

\author{J.~Zejma}
\affiliation{Marian Smoluchowski Institute of Physics, Jagiellonian University, 30-348 Cracow, Poland}

\author{N.~Ziehl}
\affiliation{ETH Z\"{u}rich, Institute for Particle Physics and Astrophysics, CH-8093 Z\"{u}rich, Switzerland}

\author{G.~Zsigmond}	
\affiliation{Paul Scherrer Institut, CH-5232 Villigen PSI, Switzerland}

\begin{abstract}
We report on a search for a new, short-range, spin-dependent interaction using a modified version of  the experimental apparatus used to measure the permanent neutron electric dipole moment at the Paul Scherrer Institute. This interaction, which could be mediated by axion-like particles, concerned the unpolarized nucleons (protons and neutrons) near the material surfaces of the apparatus and polarized ultracold neutrons stored in vacuum. 
%To increase the sensitivity to this interaction, one aluminum electrode was replaced by an electrode made of copper. 
% we searched for a hypothetical short-range, spin-dependent interaction, which could be mediated by axion-like particles, between nucleons on the surface of the electrodes and the stored ultracold neutrons. 
%With an optimized method of combing online and offline magnetic-field measurements using an array of cesium magnetometers and a dedicated mapper, respectively, developed from Monte-Carlo methods using synthesized data, the dominant systematic effect owing to the presence of a magnetic-field gradient over the storage chamber was reduced, thanks to an unprecedentedly low uncertainty reached.
The dominant systematic uncertainty resulting from magnetic-field gradients was controlled to an unprecedented level of approximately $\SI{4}{pT/cm}$ using an array of optically-pumped cesium vapor magnetometers and magnetic-field maps independently recorded using a dedicated measurement device. 
%The dominant systematic uncertainty resulting from magnetic-field gradients was controlled to an unprecedented level by combining magnetic-field measurements of 15 stationary optically-pumped cesium vapor magnetometers and magnetic-field maps previously recorded using a fluxgate sensor. 	
No signature of a theoretically predicted new interaction was found, and we set a new limit on the product of the scalar and the pseudoscalar couplings $g_sg_p\lambda^2 < 8.3 \times 10^{-28}\,\si{\meter\squared}$ (95\% C.L.) in a range of $\SI{5}{\micro\meter} < \lambda < \SI{25}{\milli\meter}$ for the monopole-dipole interaction. 
This new result confirms and improves our previous limit by a factor of 2.7 and provides the current tightest limit obtained with free neutrons. 
%Using the upgraded {\color{blue} new} instrument, n2EDM, for a refined search {\color{blue} of the neutron electric dipole moment,} we expect an improvement in sensitivity by two orders of magnitude.
\end{abstract}

 %magnetic resonance spectroscopy

\maketitle

%%%%%%%%%%%%%%%%%%%%%%%%%%%%%%%%%%%%%%%%%%%%%%%%%%%%%%%%%%%%%%%%%%%%%%%%%%%
\section{Introduction}
The extremely successful Standard Model (SM) of particle physics provides testable experimental predictions usually agreeing with laboratory measurements and astronomical observations at the highest levels of accuracy \cite{Tanabashi2018,Zyla2020}. It is therefore considered as the best theory to describe the fundamental building blocks of the Universe at current measurement sensitivities. However, together with the Cosmological Standard Model, it leaves some phenomena unexplained, e.g., the observed matter-antimatter imbalance also known as baryon asymmetry of the Universe~(BAU) or the nature of dark matter (DM) and dark energy. Therefore, new physics beyond the SM~(BSM) are needed.  
Searches for permanent electric dipole moments~(EDM)~\cite{Graner2016,ACME2018,Abel2020}, which could provide evidence for BSM {\it CP} violation, are playing a vital role in constraining theoretical models and eventually explaining the BAU~\cite{Morrissey2012}. Nevertheless, nonobservation of an EDM constrains the {\it CP}-violating phase in the strong interaction to a value that is particularly small ($\bar{\theta} < \num{e-10}$, where $\bar{\theta}$ can take any value between 0 and $2\pi$)~\cite{Zyla2020},
%ten orders of magnitude smaller than theoretical expectation
constituting another big puzzle known as the strong {\it CP} problem~\cite{Peccei1977, Weinberg1978, Wilczek1978}.  

Although the {\it CP}-violating phase of the weak interaction in the SM is not small, it is insufficient to explain the BAU~\cite{Zyla2020}. 
%Furthermore, new physics is needed to describe the nature of DM and dark energy at the level of fundamental building blocks and interactions.
In contrast, many BSM theories predict the existence of new particles. In general, these theories can be categorized into two broad sectors, i.e., ultraviolet~(UV) and infrared~(IR) modifications of the SM\@. 
In the UV sector, new particles predicted by super-symmetric theories~\cite{Zyla2020} are expected to be heavy with masses on the order of or larger than \SI{100}{GeV}. They have been extensively searched for at the Large Hadron Collider~(LHC) in the past decade, so far without a signal~\cite{Butler2017,Masetti2018}. 
Theories predicting an extension in the IR regime 
%{\color{blue} \sout{at energies below the electron-volt scale. 
%These models predict}
anticipate particles with very low masses, often referred to as weakly interacting sub-eV/slim particles~(WISPs)~\cite{Jaeckel2010}, which couple very weakly to visible matter. 

In 1984, Moody and Wilczek~\cite{Moody1984} proposed to search for a new, short-range, spin-dependent (SRSD) interaction, which could be mediated by very light, weakly coupled, spin-0 bosons being well motivated candidates of WISPs. 
For spin-0 bosons, only two options exist to couple to fermions, either via a scalar or a pseudoscalar vertex with the coupling constants $g_s$ and $g_p$, respectively.
For a fermion-fermion interaction with only one boson exchange, the scalar and the pseudoscalar vertices permit three distinct interactions in a $\left({\rm monopole}\right)^2$, $\left({\rm dipole}\right)^2$, or monopole-dipole virtual boson fields, involving $g_s\!^2$, $g_p\!^2$, or $g_s g_p$, respectively. One prominent candidate for the mediator particle of these $\left({\rm monopole}\right)^2$, $\left({\rm dipole}\right)^2$, and monopole-dipole interactions is the axion, which is the pseudo-Goldstone boson~\cite{Weinberg1978} arising from the spontaneous breaking of the Peccei-Quinn symmetry~\cite{Peccei1977} introduced in 1977 to solve the strong {\it CP} problem, and is today often referred to as the ``canonical QCD axion.''
%Another new scalar boson giving rise to a short-range, spin-dependent (SRSD) interaction, {\color{blue} \sout{called the axion}}, was proposed in 1984 by Moody and Wilczek~\cite{Moody1984}.
In general, the mediator particle of these  interactions need not be the canonical QCD axion~\cite{Kim1979,Kim2010}, but may be other hypothetical bosons. These might be spin-0 axion-like particles (ALPs), which have similar properties to the canonical QCD axion, or very light spin-1 bosons~\cite{Fayet1990, Fayet1996} coupling via the vector ($g_v$) and the axial-vector ($g_A$) vertices. 
Many experiments world wide~\cite{Budker2014,Guigue2015,Rybka2015PRD,Abel2017,Zhong2018PRD,Alesini2019PRD,Brun2019,Garcon2019,Ouellet2019PRL,Lee2020PRL,Braine2020PRL,Crescini2020PRL} are actively searching for these particles, which are considered promising candidates as microscopic constituents of DM~\cite{Sikivie2021RvMP}.

\subsection{The monopole-dipole interaction}
Among the three interactions, the monopole-dipole interaction involving $g_s g_p$ and violating {\it P} and {\it T}~symmetries as well as combined {\it CP}~symmetry, is of particular interest, as the demonstration of {\it CP} violation would provide an evidence to one of the three essential criteria to explain the BAU~\cite{Sakharov1991}. 
The potential generated by the monopole-dipole interaction between a polarized ($\dagger$) and an unpolarized particle can be written as~\cite{Moody1984,Zimmer2010}
\begin{equation}
	V(\boldsymbol{r}) = g_sg_p^\dagger\frac{ \hbar^2}{8 \pi m^\dagger } \left( {\boldsymbol{\sigma}}^\dagger \cdot \hat{\boldsymbol{r}} \right) \left( \frac{1}{r \lambda} + \frac{1}{r^2} \right) e^{-r/\lambda}, 
	\label{Eq_GsGpPotential}
\end{equation}
where $m^\dagger$ and ${\boldsymbol{\sigma}}^\dagger$ are the mass and the Pauli matrices belonging to the spin of the polarized particle, $\hat{\boldsymbol{r}}$ is the unit vector along the distance $r$ between the particles, $\lambda=\hbar/\left(m^\dagger c\right)$ is the interaction range, and $\hbar$ is the reduced Planck's constant. 
The unpolarized and the polarized particles couple to the spin-0 boson via unitless scalar and pseudoscalar coupling constants $g_s$ and $g_p^\dagger$, respectively.  

In Refs.~\cite{Zimmer2010, Serebrov2010}, it was proposed that ultracold neutrons (UCNs) can be used to search for ALPs. We searched for such an interaction with the apparatus originally built for the search for the electric dipole moment of the neutron (nEDM)~\cite{Abel2020}, and with which we also set limits for an oscillating nEDM~\cite{Abel2017} through the axion-gluon coupling and for neutron to mirror-neutron oscillations~\cite{Abel2021} at the UCN source~\cite{Bison2020, Lauss2021} of the Paul Scherrer Institute (PSI) in Switzerland. In the experiment, polarized UCNs and polarized \magHg{} atoms populated a vacuum volume between two horizontal electrodes and an insulator ring. A sketch of the experimental apparatus is shown in Fig.~\ref{fig:Apparatus}, and details about the spectrometer and the measurement procedure are described in Sec.~\ref{sec_spectrometer}. The SRSD interaction would involve the polarized UCNs stored in the vessel and unpolarized nucleons (protons and neutrons) on the electrode surfaces. The measurements were performed by comparing the Larmor precession frequencies of stored UCNs and \magHg{}~atoms, which served as a cohabiting magnetometer, in a constant magnetic field \Bo. An ALP-mediated SRSD interaction between vessel materials and trapped particles can be considered as a pseudomagnetic field \bUcn{} influencing the precession frequency of UCNs, whereas the effect on the \magHg{}~atoms is negligible as their mass is much larger~($V(\boldsymbol{r}) \propto 1/m^\dagger$ in Eq.~\eqref{Eq_GsGpPotential}). Hence, the ratio of the spin precession frequencies of UCNs and \magHg~atoms, 
\begin{equation} \label{Eq_Rratio}
	\Rr{}^{\uparrow \downarrow} = \left(\frac{\fN}{\fHg}\right)^{\uparrow \downarrow}=\left| \frac{\gN}{\gHg} \right| 
	\left( 1 \pm \frac{\bUcn}{\left|\Bo{}\right|}  \pm\frac{\Gz \left\langle z \right\rangle}{\left|\Bo\right|}+\delta_{\rm else}\right),
\end{equation}
%
%is sensitive to this interaction while correcting for other magnetic-field changes. 
%
%Changing the magnetic-field polarity will result in a shift of	
%\begin{equation} \label{Eq_Rratio}
	%\Rr{}^{\uparrow \downarrow} 
	%= \left| \frac{\gN}{\gHg} \right| 
	%\left( 1 \pm \frac{\bUcn}{\left|\Bo{}\right|}  \pm\frac{\Gz \left\langle z \right\rangle}{\left|\Bo\right|}+\delta_{\rm sys}\right),
%\end{equation}	
%%
is sensitive to this interaction while magnetic-field changes cancel and other effects corrected for. 
Here, $\gN$ and $\gHg$ are the gyromagnetic ratios of the neutron and the \magHg, respectively, \bUcn{} is the pseudomagnetic field that would be experienced by UCNs stored in the apparatus (derived in Sec.~\ref{sec_pseudomagnetic_field}), and the $+/-$ signs correspond to  the upward/downward directions of the magnetic field according to gravity.  
By measuring \Rr{} in opposite directions of \Bo{}, the magnitude of the pseudomagnetic field can be extracted 
\begin{align} \label{Eq_bUcn}
	\bUcn &= \frac{\Rr{}^{\uparrow}-\Rr{}^{\downarrow}}{\Rr{}^{\uparrow}+\Rr{}^{\downarrow}}\left|\Bo\right|,
\end{align}
and in turn the strength of the interaction $g_sg_p$ can be deduced.
The dominant systematic effect is due to the vertical center-of-mass offset $\left\langle z \right\rangle$ between the \magHg{} atoms and the UCNs in the presence of an effective magnetic-field gradient \Gz{}~\cite{Afach2015b}. Additional effects $\delta_{\rm else}$ are described in more detail below.
%The relation between \bUcn{} and $g_sg_p$ is derived in Sec.\,\ref{sec_spectrometer}. 

\subsection{Derivation of the pseudomagnetic field} \label{sec_pseudomagnetic_field}
\begin{figure}[h!]
\centering
\includegraphics[width=.5\columnwidth]{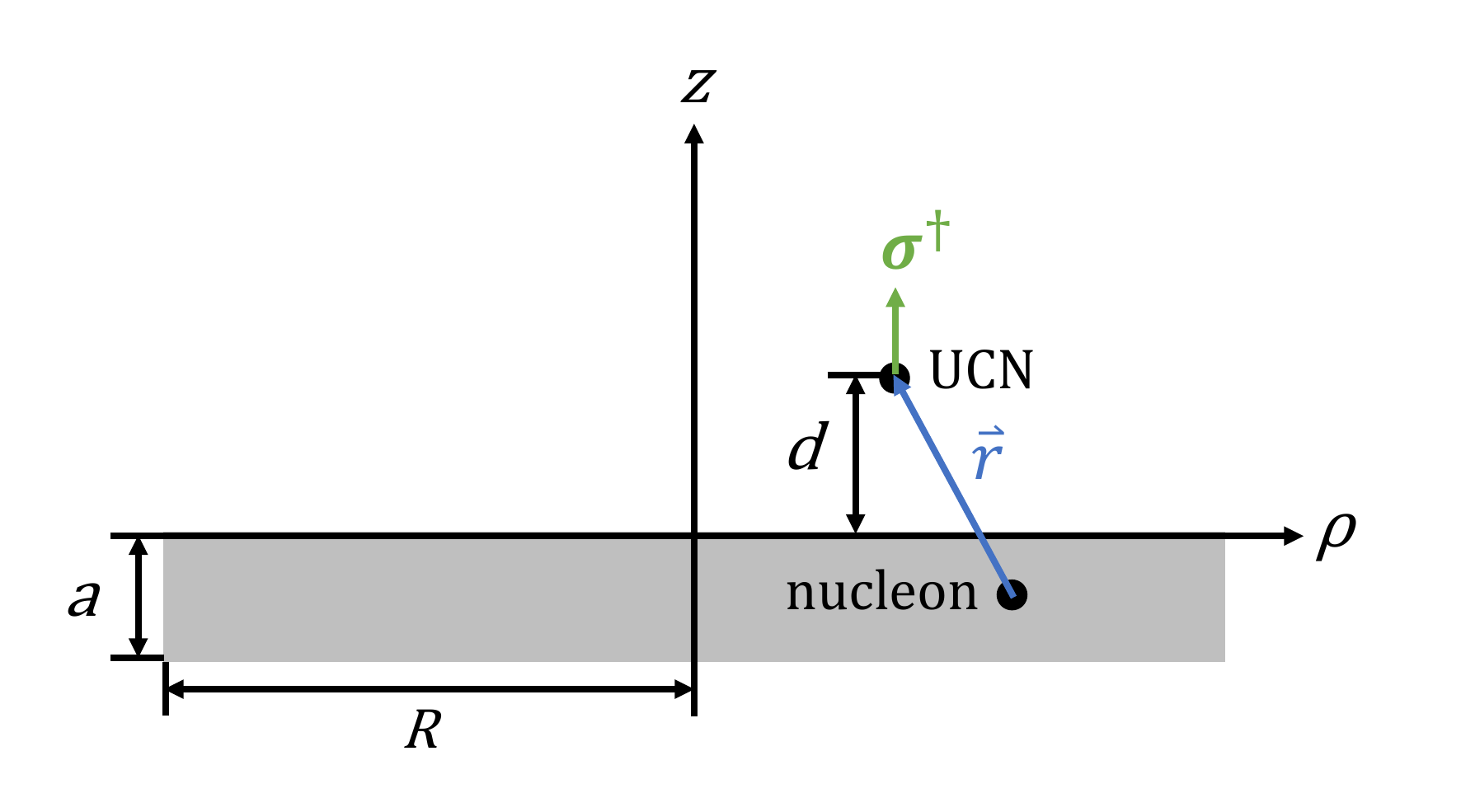}%
\caption{Schematic drawing of the interaction between one nucleon within the electrode (with a thickness $a$ and a radius $R$) and a polarized UCN in the precession chamber. The pseudomagnetic field results from integration over all nucleons in the bulk electrode.}
\label{fig:SchemeGsGpInteraction}%
\end{figure}

The interaction is described by the potential given in Eq.~\eqref{Eq_GsGpPotential}, and the effective interaction generated by one electrode in the apparatus is derived by integrating over all nucleons from the bulk matter. 
The corresponding pseudomagnetic field normal to the electrode surface at a height $d$ is written as~\cite{Afach2015}
\begin{align}
	b^{\ast}(d) 
%	&= \frac{2 V_{\rm tot}(d)}{\gamma^\dagger \hbar} \nonumber \\
	&\approx g_sg_p^\dagger \frac{\hbar N \lambda}{2 \gamma^\dagger m^\dagger} \left( 1-e^{-a/\lambda} \right) e^{-d/\lambda},
	\label{Eq_PseudoBfield}
\end{align}
where $\gamma^\dagger$ is the gyromagnetic ratio of the polarized particle, $N$ is the nucleon density depending on the material of the electrode, and $a$ is the electrode thickness, illustrated schematically in Fig.~\ref{fig:SchemeGsGpInteraction}. 

In the nEDM apparatus, both the top and the bottom electrodes made of aluminum contributed to this interaction in opposite directions, pointing from the electrodes to the UCNs stored in the chamber. We defined $z=0$ at the center of the precession chamber such that the surfaces of the top and the bottom electrodes were at $z=+H/2$ and $z=-H/2$, respectively, where $H=\SI{12}{cm}$ is the chamber height. The total pseudomagnetic field measured at a vertical coordinate $z$ can be written by summing up contributions from both electrodes using Eq.~\eqref{Eq_PseudoBfield} as 
%
%\begin{align} \label{Eq_bAlps}
%	\bAlps(z) &= g_sg_p^\dagger \frac{\hbar \lambda}{2 \gamma^\dagger m^\dagger} \left( 1-e^{-\frac{a}{\lambda}} \right) \nonumber \\
%	& \quad \times \left( \Nbot e^{-\frac{z+H/2}{\lambda}} - \Ntop e^{-\frac{H/2-z}{\lambda}} \right),
%\end{align}
\begin{equation} \label{Eq_bAlps}
	\bAlps(z) = g_sg_p^\dagger \frac{\hbar \lambda}{2 \gamma^\dagger m^\dagger} \left( 1-e^{-a/\lambda} \right) 
	\left( \Nbot e^{-\left(z+H/2\right)/\lambda} - \Ntop e^{-\left(H/2-z\right)/\lambda} \right),
\end{equation}
where \Nbot{} and \Ntop{} are the nucleon densities of the bottom and the top electrodes, respectively. 

%The pseudomagnetic field, \bAlps{}, results from the interaction between the polarized UCN and the nucleons on the electrodes. 
Since the UCNs have very low kinetic energies, their trajectories are strongly influenced by gravity. As a result, they were not uniformly distributed within the precession chamber; instead, the center of mass of the UCNs was shifted to negative $z$ values. This effectively resulted in a center-of-mass offset $\left\langle z \right\rangle=\SI{-3.9(3)}{mm}$~\cite{Abel2020} with respect to that of the cohabiting \magHg{} atoms. A linear approximation of the normalized vertical-UCN-density function is given as~\cite{Afach2015}
\begin{equation} \label{Eq_ComOffset}
	\rhoUcn{z} =\frac{1}{H}\left( 1 + \frac{12 \left\langle z \right\rangle }{H^{2}} z\right).
\end{equation} 
The setup was sensitive to interactions of short ranges, approximately from \si{\micro \meter} to \si{\mm}, a similar range as in Refs.~\cite{Guigue2015, Afach2015}, therefore, the UCN-density distribution can be simplified to a constant density at distances close to the surfaces of the electrodes, $\rhoUcn{-H/2}$ and $\rhoUcn{+H/2}$~\cite{Afach2015}. The effective pseudomagnetic field, defined as pointing upwards with respect to gravity, experienced by all UCNs within the precession chamber, is solved analytically by integrating over the chamber height 
%\begin{widetext}
\begin{equation}
\begin{aligned} \label{Eq_bUcnDef}
	\bUcn &= \int_{\frac{-H}{2}}^{\frac{+H}{2}} \bAlps(z) \rhoUcn{z} \diff{z}  \\
	&= g_sg_p^\dagger \frac{\hbar \lambda}{2 \gamma^\dagger m^\dagger} \left( 1-e^{-a/\lambda} \right) \int_{\frac{-H}{2}}^{\frac{+H}{2}} \left[ \Nbot \rhoUcn{\frac{-H}{2}} e^{-\left(z+H/2\right)/\lambda} - \Ntop \rhoUcn{\frac{+H}{2}} e^{-\left(H/2-z\right)/\lambda}\right] \diff{z}  \\ 	
	&= g_sg_p^\dagger \frac{\hbar \lambda^2 \left[ H\left(\Nbot-\Ntop\right)-6\left\langle z \right\rangle\left(\Nbot+\Ntop\right) \right]}{2 \gamma^\dagger m^\dagger H^2} \left( 1-e^{-a/\lambda} \right) \left( 1-e^{-H/\lambda} \right),	
\end{aligned}	
\end{equation}
where the top and the bottom electrodes contribute in opposite directions. 
%\end{widetext}

%\pagebreak

%%%%%%%%%%%%%%%%%%%%%%%%%%%%%%%%%%%%%%%%%%%%%%%%%%%%%%%%%%%%%%%%%%%%%%%%%%%
\section{Measurement with the \lowercase{n}EDM spectrometer} \label{sec_spectrometer}
\begin{figure}[h!]
	\centering
	\vspace{.7cm}
	\includegraphics[width=0.5\columnwidth]{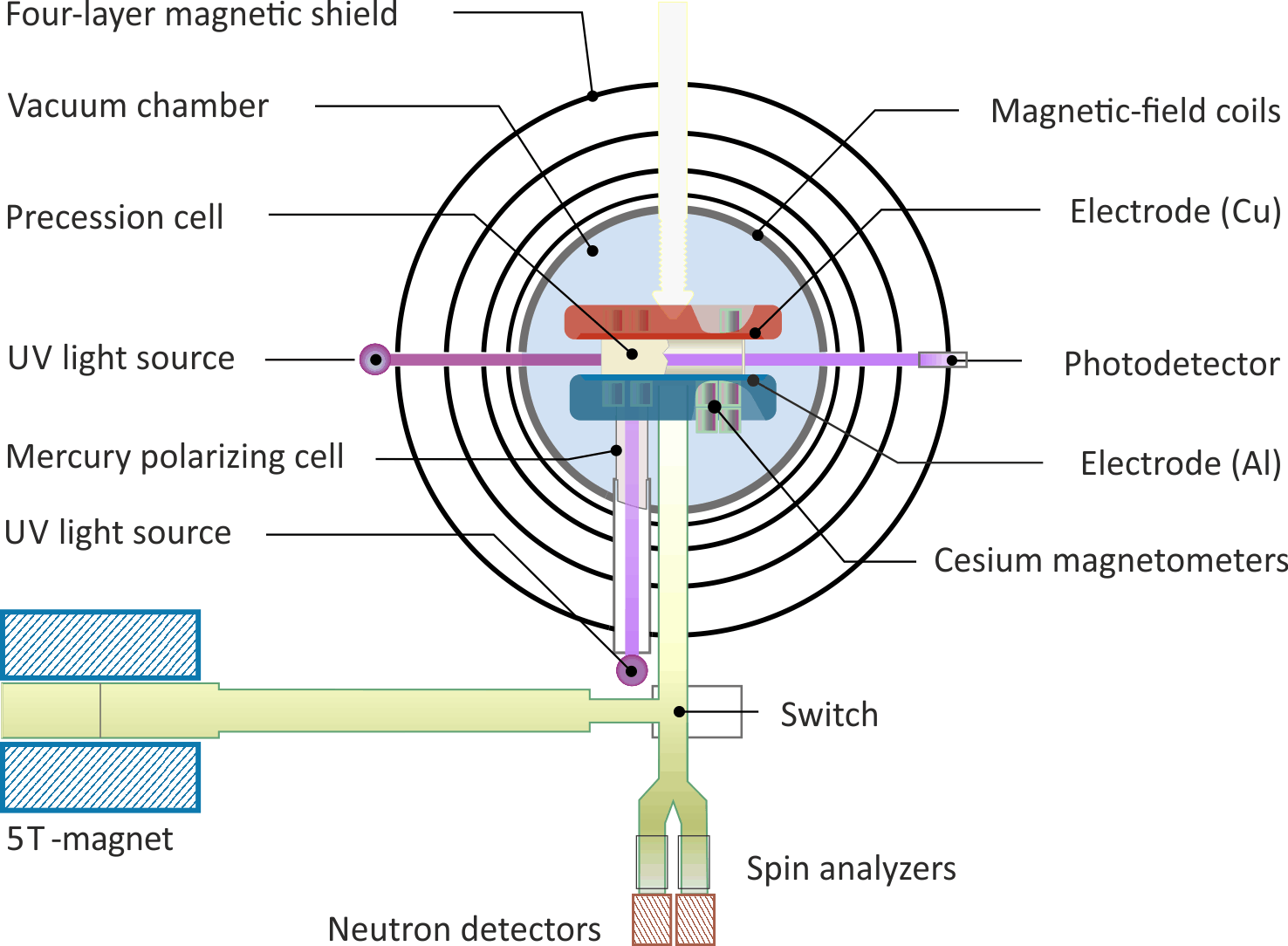}
	\caption{Schematic drawing of the modified nEDM apparatus, where the top electrode was replaced with one made of copper, used to search for a new, ALP-mediated, SRSD interaction.}
	\label{fig:Apparatus}
\end{figure}

For the measurement, we exchanged the top electrode of the nEDM spectrometer by one made out of copper to increase the nucleon density, which increased the sensitivity to this interaction. A sketch of the modified apparatus is shown in Fig.~\ref{fig:Apparatus}, where no electric field was applied during these measurements. The nucleon density\footnote{The nucleon densities were calculated using the material densities and the atomic masses obtained from MaTeck's periodic table of elements (\href{https://mateck.com/en/}{https://mateck.com/en/}, accessed 2023-02-26) and Ref.~\cite{Lide2004}.} for the top electrode was $\Ntop = N_{\rm Cu} = 5.40\times10^{30}\,\si{\per\cubic\meter}$, whereas the bottom electrode made of aluminum remained, with $\Nbot = N_{\rm Al} = 1.63\times10^{30}\,\si{\per\cubic\meter}$. 
In this way, an asymmetric pseudomagnetic field \bAlps{} was generated, increasing the sensitivity by a factor of 7.7 compared to that using both electrodes made of aluminum.

Further, we exchanged the ultraviolet light source of the probe beam of the \magHg{}-comagnetometer (HgM) from a mercury discharge lamp~\cite{Green1998} to a locked frequency quadrupled diode laser\footnote{TOPTICA Photonics AG. Product description TA / FA-FHG pro. Accessed 2022-01-11. \href{http://www.toptica.com/products/tunable-diode-lasers/frequency-converted-lasers/ta-fhg-pro/}{http://www.toptica.com/products/tunable-diode-lasers/frequency-converted-lasers/ta-fhg-pro/}}  with a wavelength of \SI{253.7}{nm}~\cite{Ban2018} to maximize the sensitivity of the HgM readout. 
The rest of the apparatus remained unchanged compared to our previous search for an SRSD interaction mediated by an ALP~\cite{Afach2015}. Polarized mercury vapor and polarized UCNs precessed in a cylindrical storage chamber of $H=\SI{12}{cm}$ height. Its side walls were made of normal polystyrene with a radius of $R=\SI{23.5}{cm}$, and the inside was coated with deuterated polystyrene~\cite{Bodek2008}. The bottom was closed off by an aluminum electrode with a central shutter for UCNs and a smaller shutter for mercury vapor.
%While the lower one remained unchanged made of aluminum, the upper one was replaced by one made of copper to increase the nucleon density. 
All inner metal surfaces of the storage cylinder, including the aluminum and the copper electrode surfaces, were coated with a thin layer ($\sim 1\text{--}\SI{3}{um}$ thickness) of diamond-like carbon~\cite{Atchison2006,Atchison2007} to improve the coherence and storage times of UCNs and \magHg{} atoms. 
Additionally, a total of 15 cesium magnetometers~(CsM), of which seven were installed above and eight below the precession chamber, were used to monitor the magnetic-field gradient \Gz{} along the chamber axis~\cite{Abel2020a}. 
A cosine-theta coil comprising around 50 turns powered with a current of about \SI{17}{mA} created a stable and uniform magnetic field of $\left|\Bo\right| \approx \pm \SI{1036}{nT}$ vertically across the chamber. 
%with a stability of $\Delta B/B_0 \leq\num{e-3}$. 
Additionally, 30 trimcoils were installed around the vacuum chamber that could be used to create a certain magnetic-field configuration if required. Four layers of cylindrical mu-metal shield and a surrounding field compensation coil system~\cite{Afach2014a} were used to passively and actively improve the stability of the magnetic field.

\begin{figure}[h!]
\centering
\includegraphics[width=0.5\columnwidth]{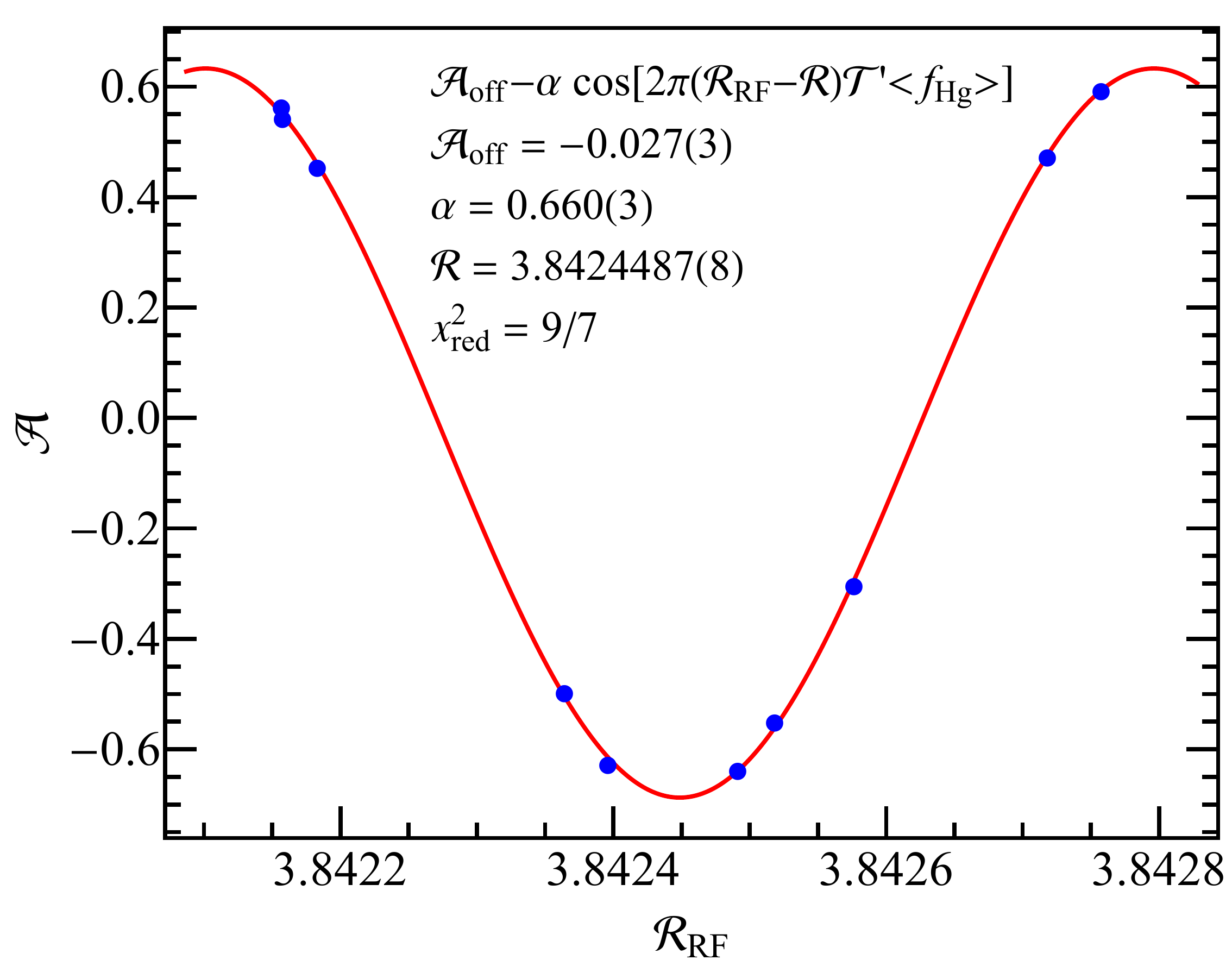}%
\caption{Ramsey pattern: The asymmetry $\mathcal{A}$ is plotted as a function of the frequency ratio $\mathcal{R}_{\rm RF}$. Blue markers are data from an ALP measurement run (run 012951), and the red line is the fit to Eq.~\eqref{Eq_RamseyPattern}. The uncertainties on $\mathcal{A}$ are smaller than the marker size. The resonant-frequency ratio $\mathcal{R}$ is at minimal $\mathcal{A}$. 
%Green diamonds indicate the working points used for nEDM data taking~\cite{Abel2020}, which were not applied in ALP measurements.
}
\label{fig:RamseyFringe}%
\end{figure}

%The measurements were performed by confining UCN in the precession chamber. 
In each measurement {\it cycle}, polarized UCNs in the magnetic field \Bo{} were used for Ramsey's method of separated oscillatory fields~\cite{Ramsey1950}. A cycle started by bringing the polarized UCNs into the precessing chamber, followed by the filling of the polarized \magHg{} atoms. 
%Simultaneously, polarized \magHg{} atoms stored in the same volume as the UCNs precessed freely under the same holding magnetic field~\cite{Green1998}. 
When both particle species were prepared in their initial stages in the storage chamber, two low-frequency pulses were applied consecutively to the \magHg{} atoms and to the UCNs to flip their spins to the transverse plane. These pulses are called $\pi/2$-pulses. After a free-spin-precession duration of $\Tp{}=\SI{180}{s}$, a second $\pi/2$-pulse, in phase with the first one, was applied to the UCNs to further tip there spins for another $\pi/2$. Afterwards, a spin-sensitive detection system~\cite{Afach2015e,Ban2016} counted neutrons in spin-up~(\Nup{}) and spin-down~(\Ndo{}) states at the end of the cycle, from which the asymmetry $\mathcal{A} = \left(\Nup-\Ndo\right)/\left(\Nup+\Ndo\right)$ was calculated. The HgM precession frequency was measured using a circularly polarized, resonant ultraviolet laser beam at \SI{253.7}{nm} wavelength, traversing the chamber while recording the spin-precession-modulated light intensity with a photo-multiplier tube~\cite{Ban2018}.
A measurement {\it run} consisted of approximately ten cycles with different spin-flipping frequencies $f_{\rm n,RF}$. These frequencies that were applied as $\pi/2$-pulses were slightly detuned from the resonant \fN{}. 
The scan of $f_{\rm n,RF}$ throughout every run led to different asymmetries of the final-state neutrons in each cycle. 
In combination with the measured HgM frequency of a single cycle, an interference pattern, {\it Ramsey pattern}, may be plotted by displaying the asymmetry as a function of the frequency ratio $\Rr{\rm RF} = f_{\rm n,RF} / \fHg$~(Fig.~\ref{fig:RamseyFringe}).

The central Ramsey fringe, approximated well with a cosine function, 
\begin{equation} \label{Eq_RamseyPattern}
	\mathcal{A} = \mathcal{A}_{\rm off} - \alpha \cos \left[ 2 \pi \left( \Rr{\rm RF} - \Rr{}\right) \Tp{}' \left\langle \fHg \right\rangle \right],
\end{equation}
was fitted to the interference pattern, where $\left\langle \fHg \right\rangle$ is the average HgM frequency of all cycles within the run. Three parameters, the asymmetry offset $\mathcal{A}_{\rm off}$, the visibility (or {\it Ramsey contrast}) $\alpha$, and the resonant-frequency ratio $\Rr{} = \fN{}/\fHg{}$, were extracted. $\Tp{}' = \Tp{} + 4\tau_{\rm n}/\pi$ is the effective time related to the fringe width, where $\tau_{\rm n}=\SI{2}{s}$ is the length of a neutron $\pi/2$-pulse. By taking the ratio of the two frequencies, we compensated for magnetic-field changes from one cycle to the next one, which cancel in Eq.~\eqref{Eq_Rratio}. 
%A change of the resonant \Rr{} with a change of the magnetic-field polarity is the signature of an SRSD interaction between polarized UCN within the precession chamber and un-polarized nucleons of the electrodes. 
%

%\subsection{Measurement strategy}
The difference of resonant-frequency ratios \Rr{}, taken for direction inverted magnetic-field configurations permits the extraction of \bUcn{} using Eq.~\eqref{Eq_bUcn}.
%By comparing precession-frequency ratio \Rr{} for inverted magnetic field configurations we extract \bUcn{} using equation~\eqref{Eq_bUcn}. 
As \Rr{} is a linear function of \Gz{} (Eq.~\eqref{Eq_Rratio}), we intentionally took data at different vertical magnetic-field gradients. Each measurement run thus comprised a certain magnetic-field configuration. 
%
% A measurement for a given magnetic-field gradient consisted of several cycles. 
The vertical magnetic-field gradient was generated by applying dedicated currents in a pair of trim coils installed above and below the vacuum tank~\cite{Afach2014}. 
In addition to the high-granularity maps of the magnetic field, taken using a dedicated measurement device~\cite{Abel2022}, the spatial distribution of the magnetic field was measured continuously with a sampling rate of \SI{1}{Hz} with 15 CsM\@.
%~\footnote{One CsM was malfunctioning and not used in the analysis.}. 
%Additionally, as part of the nEDM measurement a  inside the precession chamber was taken using a dedicated measurement device\,\cite{Abel2022}. Section~\ref{subsec_gz} describes how the high resolution magnetic-field maps and the data from the 15 CsM were efficiently combined to provide an accurate estimate of \Gz{}. 
%In each cycle Ramsey's method of separated oscillating fields with different $\Rr{\rm RF}$ was applied, from which the frequency ratio \Rr{} between UCN and HgM was extracted, see Fig.~\ref{fig:RamseyFringe}.
%$\Rr{}$ was measured for different \Gz{}. 
Using a linear fit of \Gz{} versus $\Rr{}\left(\Gz{}\right)$ by correcting for all known systematic effects of \Rr{} (Sec.~\ref{subsec_r}) and precisely determining \Gz{} (Sec.~\ref{subsec_gz}), $\Rr{}_0^\uparrow$ and $\Rr{}_0^\downarrow$, resonant-frequency ratios at $\Gz{} = 0$ for both \Bo{} directions, were precisely determined (Sec.~\ref{subsec_crosspoint}). 

%During this measurement, the main difference compared to ALP2015 was the replacement of the top electrode. Previously (ALP2015), both the top and the bottom electrodes were made of aluminum, whereas in this work, the top electrode was replaced with a copper plate, which had a larger nucleon density.
%The replacement of the top electrode material with copper increased the sensitivity to \bUcn{} compared to the previous work~\cite{Afach2014}.

%
%In ALP2015, eleven CsM were installed, which permitted the estimation of \Gz{} with an uncertainty of \SI{8}{pT/cm} using the jackknife method~\cite{Afach2014}. 
%Further, a total of 16 optically pumped cesium vapor magnetometers (CSM) were installed to measure the spatial distribution of the magnetic field in the vicinity of the storage chamber~\cite{Abel2020a}. During the measurement 15 CsM were working and recorded local field values with a sampling rate of \SI{1}{Hz}.

%%%%%%%%%%%%%%%%%%%%%%%%%%%%%%%%%%%%%%%%%%%%%%%%%%%%%%%%%%%%%%%%%%%%%%%%%%%
\section{Data analysis}
\subsection{Extraction of the resonant-frequency ratio \texorpdfstring{$\mathcal{R}$}{\it R}} \label{subsec_r}
For each measurement run, a resonant \Rr{} was obtained by fitting the Ramsey pattern, Eq.~\eqref{Eq_RamseyPattern}.
Various statistical uncertainties and systematic effects influenced the measurement of \Rr{}. 
Four effects were considered for each measurement cycle as stochastic uncertainties. These include the neutron counting statistics, the uncertainty of the estimated HgM frequency, the magnetic-field-gradient (\Gz{}) drift between cycles, and the Ramsey-Bloch-Siegert shift~\cite{Bloch1940,Ramsey1955} induced by the $\pi/2$-pulse of the HgM\@ onto the neutron spin. 
The last effect resulted from the fact that the circularly rotating magnetic field applied to the \magHg{}~atoms resulted in small random tilts of the neutron spins. 
Each effect resulted in an uncertainty on the measured asymmetry $\mathcal{A}$, which was further propagated to the uncertainty of \Rr{} according to Eq.~\eqref{Eq_RamseyPattern} using the fitted values for $\alpha$ and $\Rr{}$. Table~\ref{Tab_StatisticalSensitivityOfR} shows the average uncertainties of each effect for all measurement cycles. We calculated the reduced chi-square $\chi^2_{\rm red}$ from the Ramsey fit for all 17 runs including both directions of \Bo{}, and the mean value was \num{9.15}. 
Assuming pure Poisson statistics, the reduced chi-square $\chi^2_{\rm red}$ should be approximately 1. A scaling factor of 2.8, the square-root of the average $\chi^2_{\rm red}$ values excluding one run with $\chi^2_{\rm red} > 20$, was applied to the \Rr{} errors obtained from the fit of all runs to account for stochastic errors that were unaccounted for\footnote{The reason of excluding this run was due to the fact that its $\chi^2_{\rm red}$ was almost two times larger than the second largest $\chi^2_{\rm red}$ among all 17 runs. However, this run was still included in the final analysis. We verified that by excluding this specific run, the final result remained unchanged.}.
%{\color{red} How many runs had a $\chi^2>20$? Do we still include them in the analysis?}
	
\begin{table}[h!]
	\centering
	\begin{tabular}{|l|c|c|}
		\hline
		Effect / \num{1e-7} & $B_0$ up & $B_0$ down \\
		\hline
		Neutron counts & 1.84 & 2.26 \\ 
		HgM frequency & 0.75 & 0.69 \\ 
		Gradient drift & 0.02 & 0.02 \\  		
		\magHg{} spin-flip pulse & 0.07 & 0.23 \\  
		\hline 
		Total stochastic effects & 2.02 & 2.41 \\
		\hline	
	\end{tabular}
	\caption[]{Stochastic uncertainties of $\mathcal{R}$ from all measurement cycles. The total numbers of neutrons have mean values of approximately 14000 and 10000 for the magnetic field pointing upwards and downwards, respectively.}
	\label{Tab_StatisticalSensitivityOfR}
\end{table}

We recall that the dominant systematic effect is the gravitational shift $\delta_{\rm grav}$ resulting from the center-of-mass offset $\left\langle z \right\rangle$ between the UCN and the \magHg{}~ensembles (Eq.~\eqref{Eq_Rratio}). In the presence of a vertical magnetic-field gradient~\Gz{}, both species measure slightly different volume averages of the magnetic field. The gravitational shift is calculated as 
\begin{align} \label{Eq_deltaGrav}
	\delta_{\rm grav} =\frac{{\left\langle B_z \right\rangle}_{\rm n}}{{\left\langle B_z \right\rangle}_{\rm Hg}} - 1 
	= \pm\frac{\Gz \left\langle z \right\rangle}{\left|\Bo\right|}, 	
\end{align}
where 
%
%\begin{align}
%	\Gz{} &= G_{1,0} + G_{3,0} \left( \frac{3H^2}{20} - \frac{3R^2}{4} \right) \nonumber \\ 
%	& \qquad + G_{5,0} \left( \frac{5R^4}{8} - \frac{3R^2H^2}{8} + \frac{3H^4}{112} \right)
%	\label{Eq_Ggrav} 
%\end{align}  
\begin{equation}
	\Gz{} = G_{1,0} + G_{3,0} \left( \frac{3H^2}{20} - \frac{3R^2}{4} \right) 
	+ G_{5,0} \left( \frac{5R^4}{8} - \frac{3R^2H^2}{8} + \frac{3H^4}{112} \right)
	\label{Eq_Ggrav} 
\end{equation}
is the effective magnetic-field gradient parallel to the gravitational gradient calculated using a polynomial expansion of the magnetic field to fifth degree~\cite{Abel2019, Abel2022}. $G_{\ell,m}$ are expansion coefficients of degree~$l$ and order~$m$ of the harmonic polynomial, whereas $H = \SI{12}{cm}$ and $R = \SI{23.5}{cm}$ are the height and the radius of the precession chamber. The $+/-$ signs in Eq.~\eqref{Eq_deltaGrav} correspond to \Bo{} in upward/downward directions, respectively, with $\left\langle z \right\rangle < 0$. More details on this effect is described in Sec.\,\ref{subsec_gz}.

Other known effects $\delta_{\rm else}$ shown in Eq.~\eqref{Eq_Rratio} are summarized as 
\begin{equation}
	\delta_{\rm else}  = 
	\delta_{\rm T} + \delta_{\rm Earth} + \delta_{\rm light} + \delta_{\rm inc} + \delta_{\rm JNN},
	\label{Eq_SysEffectsR}
\end{equation} 
which are caused by the transverse magnetic-field components, the rotation of Earth, the UV laser beam for the HgM readout, the incoherent scattering of UCNs on the \magHg{} atoms, and magnetic-field fluctuations resulting from Johnson-Nyquist noise (JNN)~\cite{Johnson1928, Nyquist1928}, respectively. They may be divided into two categories. On the one hand, constant shifts of the UCN or HgM frequency lead to a deviation of \Rr{} from the ratio of pure gyromagnetic ratios. These include the first three effects, $\delta_{\rm T} $, $\delta_{\rm Earth}$, and $\delta_{\rm light}$. 
%, which were quantified and rectified. 
%$\delta_{\rm grav}$ was a factor used to extrapolate the \Rr{} value at the limit of zero gradient while correcting for $\delta_{\rm T} $ and $\delta_{\rm Earth}$. 
On the other hand, $\delta_{\rm inc}$ and $\delta_{\rm JNN}$ are pure stochastic effects, which do not shift the mean \Rr{} value, but result in an increase of the measurement uncertainty. There effects are shown in Sec.\,\ref{subsec_crosspoint}.  
%Individual errors were estimated and included in the final error budget (see Sec.\,\ref{subsec_crosspoint}). 
%Note that $\gHg > 0$ and $\gN < 0$ because of the opposite signs of their magnetic moments.  

The transverse shift $\delta_{\rm T}$ is a consequence of transverse components of the magnetic field $B_{\rm T}$. 
Ultracold neutrons in a magnetic field of \SI{1}{\micro\tesla} sample the field in the adiabatic regime of slow particles in a high magnetic field. The measured mean frequency  $\omega_{\rm n}=\gamma_{\rm n}\left\langle\sqrt{B_x^2 + B_y^2 + B_x^2}\right\rangle$ is proportional to the volume average of the magnetic-field modulus. 
By contrast, \magHg{} atoms in the same magnetic field fall into the nonadiabatic regime of fast particles in a low magnetic field, such that their spins precess at a mean frequency $\omega_{\rm Hg}=\gamma_{\rm Hg}\sqrt{\left\langle B_x\right\rangle^2+\left\langle B_y\right\rangle^2+\left\langle B_z\right\rangle^2}$  given by the volume average of the vector magnetic field.
In the presence of $B_{\rm T}$, this results in 
\begin{equation} \label{Eq_deltaBt2}
	\delta_{\rm T} 
	%	&= \frac{\frac{\left| \gN \right|}{2\pi}}{\frac{\gHg}{2\pi}} \frac{\frac{\left\langle B_{\rm T}\!^2 \right\rangle}{2 \left| \Bo \right|}}{\left| \Bo \right|} \\ 
	= \frac{\left\langle B_{\rm T}\!^2 \right\rangle}{2\Bo\!^2},
\end{equation}
with
\begin{equation}
	\left\langle B_{\rm T}\!^2 \right\rangle = \left\langle \Delta B_x^2 + \Delta B_y^2  \right\rangle
\end{equation}
being the mean-square transverse magnetic-field components, where
$\Delta B_x = B_x - \left\langle B_x \right\rangle$ and $ \Delta B_y = B_y - \left\langle B_y \right\rangle$. 
For each run, i.e., one magnetic-field configuration, we calculated $\left\langle B_{\rm T}\!^2 \right\rangle$ using the field maps~\cite{Abel2022} and corrected the resonant \Rr{} value obtained from the Ramsey fit (Eq.~\eqref{Eq_RamseyPattern}) by $\delta_{\rm T}$.

%$\delta_{\rm grav}$ and $\delta_{\rm T}$ are called {\it nonuniform magnetic terms}, originating from the low energies of the UCN ensemble. These terms, however, vanish in a perfectly homogeneous magnetic field. The others are considered as {\it secondary effects}, which are not directly correlated with the magnetic field.  
Effectively, given the rotation of Earth, the precession frequencies of UCNs and \magHg{} atoms are measured in a rotating frame of reference. They are a combination of the  Larmor frequency in a stationary frame and the Earth's rotational frequency, $f_{\rm Earth}=\SI{11.6}{\micro Hz}$. 
The associated shift in \Rr{} was corrected for by calculating
%
%\begin{align}
%	\delta_{\rm Earth} &= \mp \left( \frac{f_{\rm Earth}}{f_{\rm n}} + \frac{f_{\rm Earth}}{\fHg} \right) \cos \theta_{\rm PSI} \nonumber \\
%	&= \mp 1.4\times10^{-6}, 
%	\label{Eq_deltaEarth}
%\end{align}  
\begin{equation}
	\delta_{\rm Earth} = \mp \left( \frac{f_{\rm Earth}}{f_{\rm n}} + \frac{f_{\rm Earth}}{\fHg} \right) \cos \left( \theta_{\rm PSI} \right) 
	= \mp 1.4\times10^{-6}, 
	\label{Eq_deltaEarth}
\end{equation}  
where $\cos\left(\theta_{\rm PSI}\right)=0.738$ is the cosine of the angle between the \Bo{} direction and the rotational axis of the Earth, corresponding to the latitude of the PSI, and the $-/+$ signs correspond to the upward/downward directions of \Bo{}, respectively.
% All \Rr{} values were corrected for depending on the \Bo{} polarity. 

The third effect $\delta_{\rm light}$ is related to the UV laser traversing the precession chamber to read out the HgM signal. This value was not quantified during the measurement; therefore, we only estimate its effect and consider it as another contribution to the final uncertainty of \Rr{}. Details are given in Sec.\,\ref{subsec_crosspoint}.

\subsection{Determination of the vertical magnetic-field gradient \texorpdfstring{$G_{\rm grav}$}{G\_grav}} \label{subsec_gz}
%The dominant systematic effect on an \Rr{} measurement arises from the center-of-mass offset between the two particle ensembles in combination with an effective vertical magnetic-field gradient $\Gz{}$.
%\begin{align}
	%\Gz{} &= G_{1,0} + G_{3,0} \left( \frac{3H^2}{20} - \frac{3R^2}{4} \right) \nonumber \\ 
	%& \qquad + G_{5,0} \left( \frac{5R^4}{8} - \frac{3R^2H^2}{8} + \frac{3H^4}{112} \right), 
	%\label{Eq_Ggrav} 
%\end{align}
%which is described by a polynomial expansion~\cite{Abel2019, Abel2022}. $G_{l,m}$ are expansion coefficients in degree $l$ and order $m$ of harmonic polynomials, whereas $H = \SI{12}{cm}$ and $R = \SI{23.5}{cm}$ are the height and the radius of the precession chamber.

A total of 15 CsM, which were of scalar-type magnetometers, measured the magnitudes of the magnetic field above and below the precession chamber during a measurement, which was used to quantify \Gz{}. Because of the applied $\Bo\approx \SI{1}{\micro\tesla}\,\hat{e}_z$, the transverse fields of order \SI{1}{nT} were negligible compared to the vertical field; hence, the scalar fields measured by the CsM were the effective vertical-field component $B_z$. 

The magnetic fields measured by the CsM were described by a polynomial expansion~\cite{Abel2019}. The gradients $G_{\ell,m}$ were extracted by fitting the 15 field values measured by the CsM to the $z$ component of the polynomial expansion
\begin{equation}
	B_{\rm CsM}^{i} \left( \boldsymbol{r}^i \right) = \sum_{\ell,m} G_{\ell,m} \varPi_{z,\ell,m} \left( \boldsymbol{r}^i \right),
	\label{Eq_Bcsm}
\end{equation}
where $i=1,\,2,\,\ldots,\,15$ is the index of the CsM, $\boldsymbol{r}^i$ is the position of the corresponding CsM, and $\varPi_{z,\ell,m}$ is a function (or mode) expanded in harmonic polynomials of degree $l$ and order $m$ depending on $\boldsymbol{r}^i$. 

For each degree $\ell$, the order $m$ runs from $-\ell$ to $+\ell$ for $\varPi_{z,\ell,m}$, which gives $\left( 2 \ell + 1 \right)$ terms. It was observed that lower-degree fields were subjected to fluctuations; thus, they were measured online with CsM\@. Constrained by the number of 15 CsM, the highest full parametrization one can achieve with this method is of second-degree, which contains nine free parameters. The contributions of higher-degree fields were found to be more stable and reproducible. To resolve the problem of higher-degree fields that were not taken into account, we combined both online CsM measurements and offline field maps~\cite{Abel2022} to achieve an improved estimate of \Gz{}. With the field maps, the magnetic fields could be expanded up to sixth degree for all three components in $x$, $y$, and $z$. 
The expression of \Gz{} could thus be expanded to fifth degree (Eq.~\eqref{Eq_Ggrav}). Note that only odd gradients contribute to \Gz{} as the average-magnetic-field components given by even gradients cancel out for \magHg{} atoms and UCNs.

Several different methods to combine cycle-by-cycle CsM data with magnetic-field maps were investigated using synthesized magnetic-field readings, which worked as follows. We calculated field values at the CsM positions using the magnetic-field-map data and varying the coefficients $G_{\ell,m}^{\rm syn}$ randomly using a Gaussian distribution (Eq.~\eqref{eq_Glm_syn}). In addition, a uniformly distributed random offset in the range of $\pm \SI{120}{pT}$~\cite{Abel2020a} was added at each CsM location, accounting for possible sensor offsets. 
In total, 200 random fields were synthesized and analyzed using eight different methods~\cite{Chiu2021}.
The optimal method was selected by minimizing the difference between the synthesized and the fitted \Gz{} up to fifth degree using Eq.~\eqref{Eq_Ggrav},
	\begin{equation}
		\Delta \Gz = \Gz^{\rm syn} - \Gz^{\rm fit},
	\end{equation}	
referred to as the deviation. Each synthesized gradient composing $\Gz^{\rm syn}$ included a random error drawn from a Gaussian distribution, whose standard deviation was the uncertainty of the map, and was given as  
	\begin{equation}
		\begin{gathered}
			\begin{aligned}
				G_{\ell,m}^{\rm syn} &= G_{\ell,m}^{\rm map} + \delta_{G_{\ell,m}}^{\rm map}. \\   		
			\end{aligned}
		\end{gathered}
	\label{eq_Glm_syn}
	\end{equation}	
The uncertainty of the fitted gradient $\sigma_{\Gz}$ was calculated with 
	%\begin{align}
			%\sigma_{\Gz} &= \left[{\left(\sigma_{G_{1,0}^{\rm fit}}\right)}^2 \right. \nonumber \\ 
			%&\quad + {\left(\sigma_{G_{3,0}^{\rm fit}}\right)}^2 \left( \frac{3H^2}{20} - \frac{3R^2}{4} \right)^2 \nonumber \\ 
			%&\quad \left. + {\left(\sigma_{G_{5,0}^{\rm fit}}\right)}^2 \left( \frac{5R^4}{8} - \frac{3R^2H^2}{8} + \frac{3H^4}{112} \right)^2 \right]^{1/2},
	%\end{align}
error propagation from each gradient-fit error $\sigma_{G_{1,0}^{\rm fit}}$, $\sigma_{G_{3,0}^{\rm fit}}$, and $\sigma_{G_{5,0}^{\rm fit}}$ in Eq.~\eqref{Eq_Ggrav}. 

We concluded that the optimal method was achieved by removing the fields described by higher-degree harmonic polynomials with $\ell=3,\ldots,6$ using the map gradients and performing a second-degree fit including nine gradients up to $\ell=2$ to the residual fields.
The optimal fit method, even in the presence of CsM offsets with a standard deviation as large as $\sigma_{B_{\rm offset}} = \SI{115}{pT}$, estimated the coefficients with a deviation in the range of $\left|\Delta\Gz{}\right| \sim 2\text{--}3\,\si{pT/cm}$, and with fit uncertainties of $\sigma_{\Gz} < \SI{3.8}{pT/cm}$ (upward \Bo{}) and  $\sigma_{\Gz} < \SI{4.5}{pT/cm}$ (downward \Bo{}).

This method was applied to all 17 runs of data, consisting of a total of 170 cycles. 
The residuals $\Delta B^i = B_{\rm low}^i - B_{\rm fit}^i$ for each CsM $i$ in each cycle was calculated, where $B_{\rm low}^i$ are the field values used in the fit to the polynomial expansion up to second degree after removing higher-degree contributions, and $B_{\rm fit}^i$ are the calculated value at each CsM position using the fitted  gradients. 
All $\Delta B^i$ were $< \SI{250}{pT}$, which were below the uncertainties of the field maps. 
For each cycle, \Gz{} was calculated using the expansion up to fifth degree, where $G_{1,0}$ was obtained from the second-degree-polynomial fit, and $G_{3,0}$ as well as $G_{5,0}$ were taken from the map values. The average value of the estimated \Gz{} from all cycles in a run was taken as the vertical gradient of this magnetic-field configuration.
%For each run with the same magnetic-field configuration, . 

\begin{figure}[h!]
	\centering
	\subfloat[\label{Fig_AlphaVsGz5Up}]{\includegraphics[width=0.48\columnwidth]{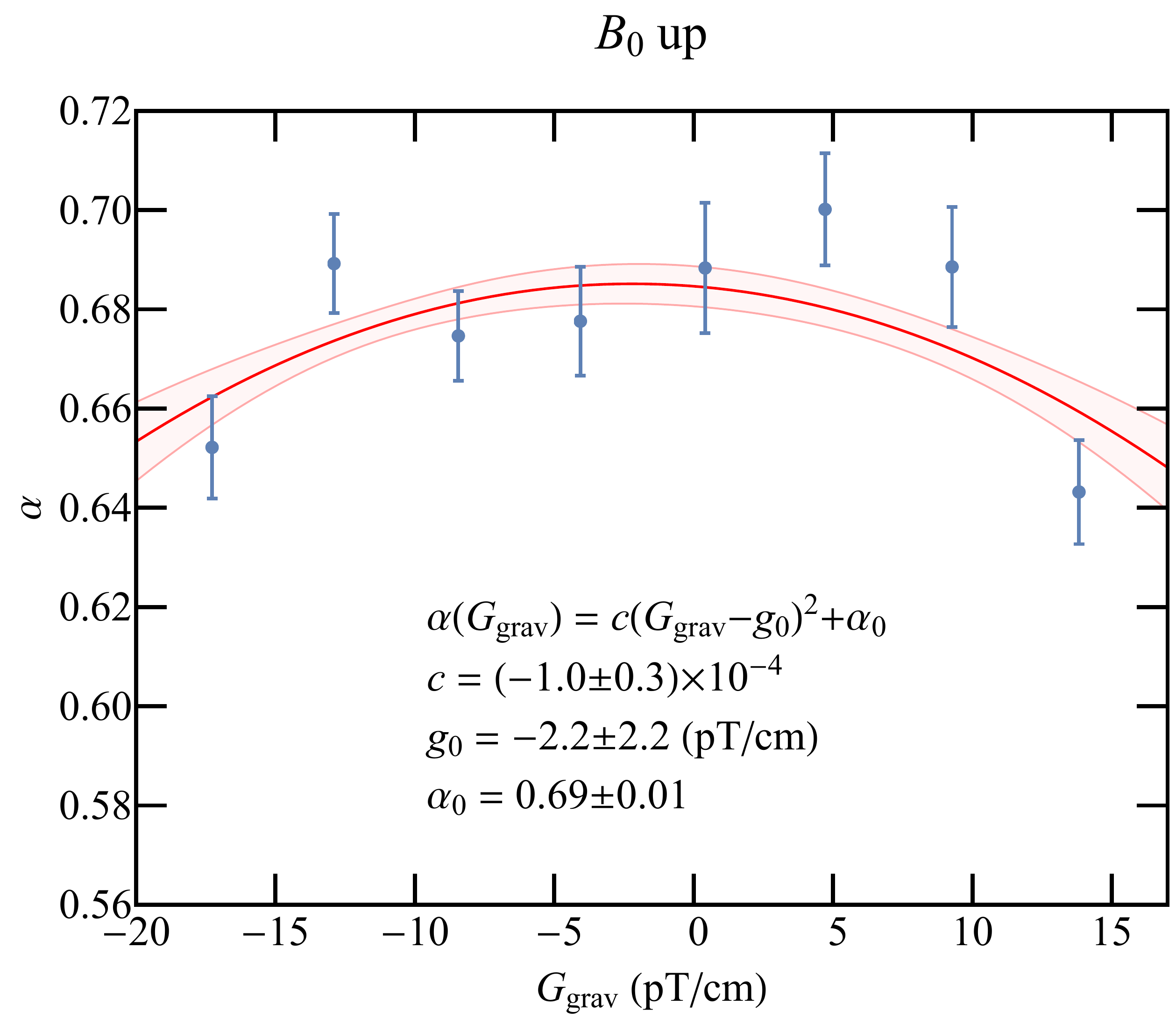}}
	\hfill
	\subfloat[\label{Fig_AlphaVsGz5Down}]{\includegraphics[width=0.48\columnwidth]{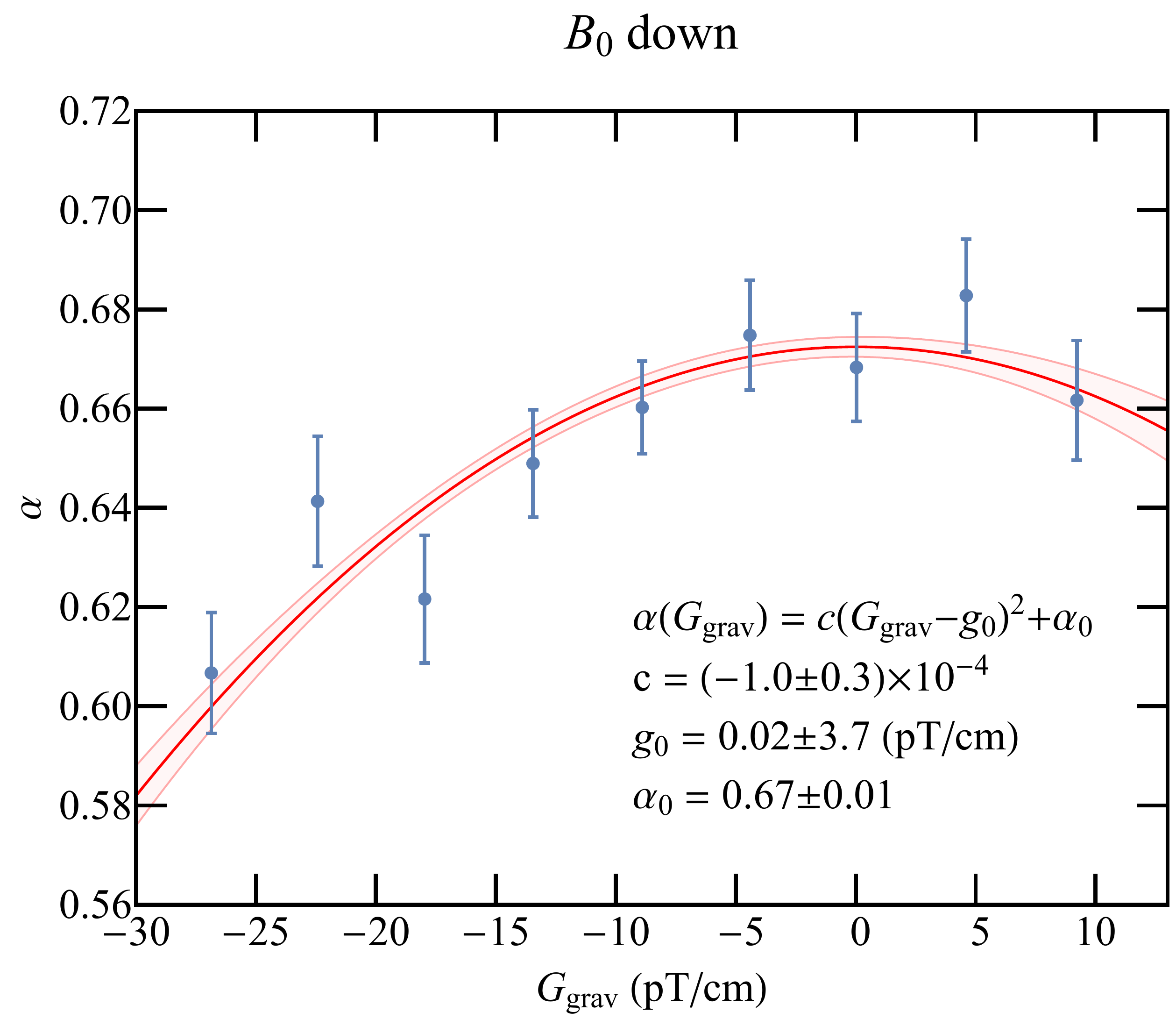}}
	\caption[]{Visibility parabola: Visibility $\alpha$ as a function of vertical magnetic-field gradient \Gz{} for \Bo{} pointing (a) upwards and (b) downwards from all 17 runs of data. The error bar on each data point is the $1\sigma$ error from the Ramsey fit scaled with a factor of 2.8. The red line and the shaded band show the fit and the 68\% confidence interval expectation. Both parabolas were fitted simultaneously with a same $c$ parameter.}
	\label{Fig_AlphaVsGz}
\end{figure}

To correct for potential systematic effects on the calculated effective gradient \Gz{}, we made use of the {\it visibility parabola}, which is the visibility of the Ramsey fringe $\alpha$ plotted as a function of \Gz{}. The parabola reaches its maximum at the minimum vertical magnetic-field gradient, where gravitationally enhanced depolarization~\cite{Harris2014, Afach2015a} is negligible. Figure \ref{Fig_AlphaVsGz} shows the parabola for \Bo{} pointing upwards (\ref{Fig_AlphaVsGz5Up}) and downwards (\ref{Fig_AlphaVsGz5Down}) with both reaching a similar maximal visibility. The parabolas were fitted with a simple parabolic function $\alpha\left(\Gz{}\right)=c(\Gz{}-g_0)^2+\alpha_0$, where $g_0$ is the expected zero gradient. The maximal visibilities were reached at $g_0^{\uparrow} = -2.2 \pm 2.2~\si{pT/cm}$ and $g_0^{\downarrow} = 0.02 \pm 3.7~\si{pT/cm}$ for the upward and the downward \Bo{} directions, respectively. The uncertainties on the fitted parameters were estimated by scaling $\chi^2_{\rm red} = \chi^2/\text{d.o.f.}$ to 1 in each parabola fit. To account for this potential shift, we corrected the effective \Gz{} of each run with $g_0^{\uparrow} = \SI{-2.2}{pT/cm}$ or $g_0^{\downarrow} = \SI{0.02}{pT/cm}$. 

%An independent cross-check of the calculated effective gradient \Gz{} was performed by analyzing the visibility of the Ramsey fringe $\alpha$ as a function of \Gz{} for \Bup{} (\ref{Fig_AlphaVsGz5Up}) and \Bdo{} (\ref{Fig_AlphaVsGz5Down}), see fig.~\ref{Fig_AlphaVsGz}.  Both \Bo{} polarities had a similar maximal visibility. The parabolas were fitted with a simple parabolic function $c(\Gz{}-g_0)^2+\alpha_0$, where $g_0$ is the expected zero gradient. The maximal visibilities is at $g_0^{\uparrow} = -1.5 \pm 2.1~\si{pT/cm}$ (\Bup{}) and $g_0^{\downarrow} = 3.8 \pm 7.7~\si{pT/cm}$ (\Bdo{}), supporting the method used for extraction of \Gz{}. 

%\FloatBarrier
\subsection{Crossing-point analysis} \label{subsec_crosspoint}
\begin{figure}[h!]
	\centering
	\includegraphics[width=0.5\columnwidth]{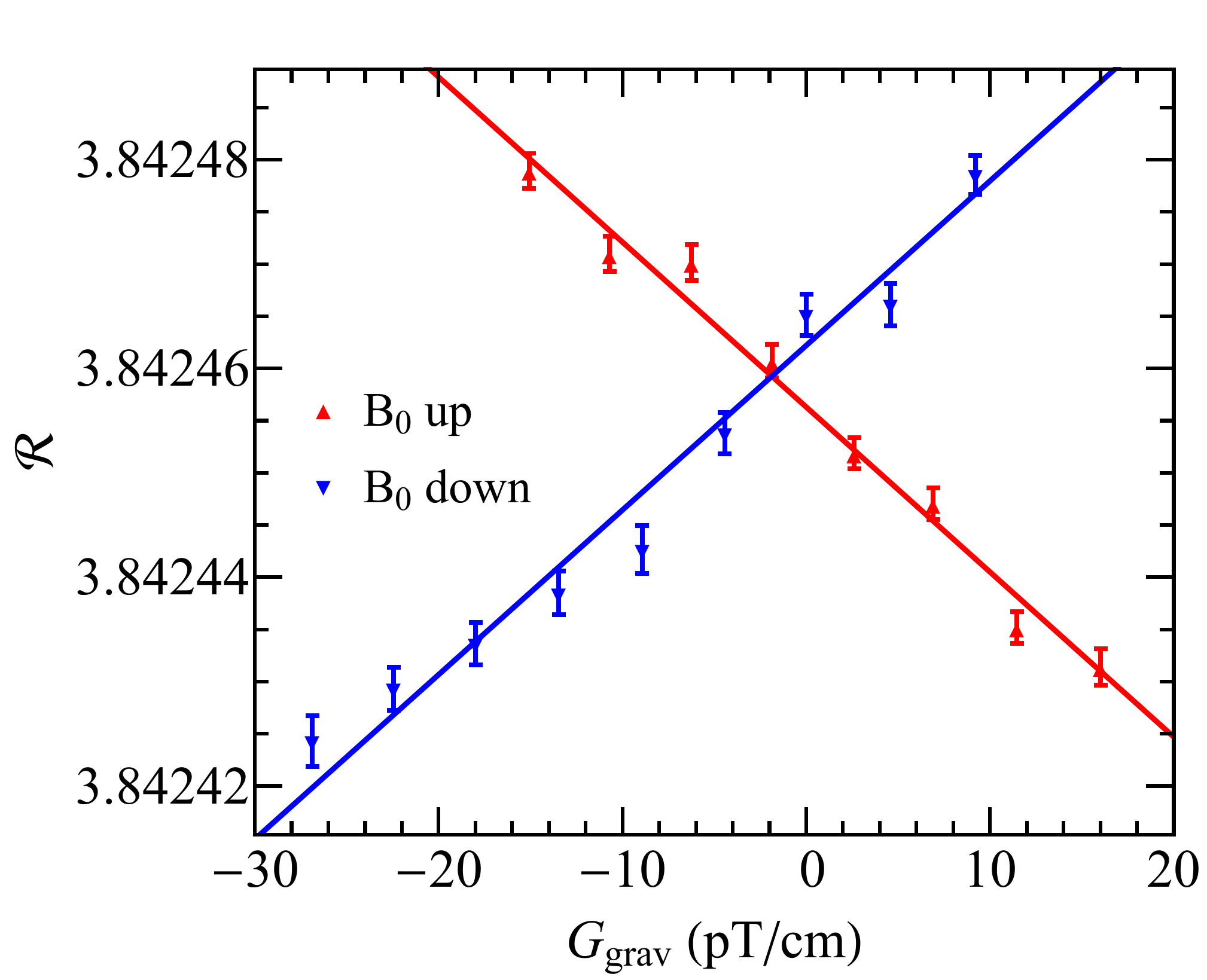}
	\caption[]{Frequency ratio \Rr{} as a function of vertical magnetic-field gradient \Gz{} for all runs with different field configurations. Red upward and blue downward triangles correspond to \Bo{} pointing upwards and downwards, respectively. The error bar for each \Rr{} is the fit error from the Ramsey fit scaled by 2.8. The error of \Gz{} for each data point with an average value of \SI{4.05}{pT/cm} over all 17 runs was not included in the error of \Rr{} individually. Instead, we considered this effect a global systematic uncertainty to all runs that resulted in an error contribution to both of the fitted $\Rr{}^{\uparrow/\downarrow}$ values. For this, the error obtained from each fit of the visibility parabolas was used (vertical magnetic-field gradient in Tab.~\ref{Tab_ErrorBudgetOnR}). Straight lines shown here were obtained using the best-fit values (Eq.~\eqref{Eq_CrossPointFit}) in the fit equations (Eq.~\eqref{Eq_Rfit}).}
	\label{Fig_CrossingPoint}
\end{figure}

%The crossing-point analysis combines the results obtained from both directions of \Bo{}.
%For this purpose, we corrected all \Rr{} values for systematic shifts $\delta_{\rm T} $ and $\delta_{\rm Earth}$. 
Figure \ref{Fig_CrossingPoint} shows the \Rr{} values, after $\delta_{\rm T} $ and $\delta_{\rm Earth}$ corrections, as a function of the corrected \Gz{} (Fig.~\ref{Fig_AlphaVsGz}). Red upward triangles and blue downward triangles are runs with \Bo{} pointing upwards and downwards, respectively.
A linear fit to the data from all runs with both directions of \Bo{} ($+/-$ correspond to upward/downward) was applied to 
\begin{equation} \label{Eq_Rfit}
	\Rr{}^{\uparrow/\downarrow} = \Rr{0}^{\uparrow/\downarrow} \left( 1\pm\frac{\left\langle z \right\rangle}{\left|\Bo{}\right|} \Gz^{\uparrow} \right)
\end{equation}
%and
%\begin{equation} \label{Eq_Rfit_down}
	%\Rr{}^{\downarrow} = \Rr{0}^{\downarrow} \left( 1-\frac{\left\langle z \right\rangle}{\left|\Bo{}\right|} \Gz^{\downarrow} \right)
%\end{equation}
simultaneously, sharing the parameter $\left\langle z\right\rangle$, while the $\Rr{0}^{\uparrow/\downarrow}$ values were kept separate for both directions. This is the so-called crossing-point analysis. 
The best fit was obtained for 
\begin{equation} \label{Eq_CrossPointFit}
	\begin{aligned}
		\Rr{0}^{\uparrow} &= 3.8424563(08), \\
		\Rr{0}^{\downarrow} &= 3.8424622(12),~\text{and} \\
		\left\langle z \right\rangle &= \SI{-0.43(2)}{cm},
	\end{aligned}	
\end{equation}
with $\chi^2_{\rm red} = \chi^2/\text{d.o.f.} = 31.9/14$. 
%{\color{red} \sout{The statistical errors on $\Rr{0}^{\uparrow/\downarrow}$ and $\left\langle z \right\rangle$ were scaled by a factor 2.8 to accommodate for unknown stochastic systematic effects.}}
The underestimated uncertainties caused $\chi^2_{\rm red}$ to be larger than 1. The uncertainties shown in Eq.~\eqref{Eq_CrossPointFit} were corrected for this stochastic error. Compared to the total statistical errors shown in Tab.\,\ref{Tab_StatisticalSensitivityOfR} summing from all known effects, these are a factor of 4--5 larger, corresponding to our initial scaling factor of 2.8 multiplied by the correction factor $\sqrt{31.9/14}$. 
The center-of-mass offset $\left\langle z \right\rangle$ was in agreement with the values found in Ref.~\cite{Abel2020}, $\left\langle z \right\rangle = \SI{-0.39(3)}{cm}$,  and Ref.~\cite{Abel2019}, $\left\langle z \right\rangle = \SI{-0.38(3)}{cm}$,
%In the absence of \bUcn{}, $\Rr{}^{\uparrow}$ and $\Rr{}^{\downarrow}$ possess the same value at $\Gz{} = 0$ (see Eq.~\eqref{Eq_RratioBucn}); thus, the two lines will cross at the limit of zero gradient if \bUcn{} does not exist. 
and the crossing point was at $G_\times=-1.9(5)\,\si{pT/cm}$.
% is in agreement with $g_0^{\uparrow/\downarrow}$ from the visibility study.  
%In addition, it is within $1\sigma$ errors of the estimated gradient uncertainties $\sigma_{\Gz{}}^{\uparrow/\downarrow}$ (see below); hence, $G_\times$ also serves as a cross-check for $\Gz{}$.

%Recall Eq.~\eqref{Eq_SysEffectsR}, systematic effects resulted in random fluctuations, which led to uncertainties on \Rr{}. There is no indication that these effects change between different runs; therefore, they were considered as global effects that influenced all runs in the same way.
In the following paragraphs, we distinguish two different kinds of uncertainties associated with \Rr{} as shown in Eq.~\eqref{Eq_Rratio}.
%Eqs.~\eqref{Eq_deltaGrav} and \eqref{Eq_SysEffectsR}. 
The first kind was systematic effects, leading to a bias, whereas the second kind was considered purely stochastic, only increasing the measurement uncertainty.

The most important systematic effect of the first kind is the shift $\delta_{\rm grav}$ induced by a vertical magnetic-field gradient. 
%Errors on the fitted \Gz{} are propagated to errors on \Rr{}. {\color{red} how?}
From the visibility parabolas (Fig.~\ref{Fig_AlphaVsGz}), $g_0^{\uparrow/\downarrow}$ were considered as systematic shifts on \Rr{}. For all runs, \Gz{} were corrected for with $g_0^{\uparrow/\downarrow}$, and the uncertainties of the fitted $g_0^{\uparrow/\downarrow}$ lead to  
\begin{equation}
	\begin{aligned}
		\sigma_{R_{\rm grav}}^{\uparrow} &= 35 \times10^{-7}~\text{and} \\
		\sigma_{R_{\rm grav}}^{\downarrow} &= 59 \times10^{-7}.
		\label{eq:Rgrav}
	\end{aligned}	
\end{equation} 

The second largest systematic effect $\delta_{\rm T}$ arises from the residual transverse field. In addition to the shift in \Rr{}, corrected for run-by-run, an error on the mean-squared transverse field $\sigma_{\left\langle B_{\rm T}^2 \right\rangle}$ was calculated making use of the concept of {\it reproducibility} of the field maps~\cite{Abel2022}, quantifying the spread in measurements of identical magnetic-field configurations taken over several years. This results in shifts of
\begin{equation}
	\begin{aligned}
		\Rr{}_{\rm T}^{\uparrow} &=\left(7.3\pm4.7\right)\times10^{-7}~\text{and} \\ \Rr{}_{\rm T}^{\downarrow} &=\left(6.4\pm4.1\right)\times10^{-7}.
		\label{eq:RT}
	\end{aligned} 
\end{equation}

 %uncertainties for each \Bo{} polarity gave $\sigma_{\left\langle B_{\rm T}^2 \right\rangle}^{\uparrow} = \SI{0.26}{nT^2}$ and $\sigma_{\left\langle B_{\rm T}^2 \right\rangle}^{\downarrow} = \SI{0.23}{nT^2}$. These translate to errors on \Rr{} of $\sigma_{R_{\rm T}}^{\uparrow} = 4.66 \times 10^{-7}$ and $\sigma_{R_{\rm T}}^{\downarrow} = 4.10 \times 10^{-7}$. 

The third systematic shift $\delta_{\rm light}$ may occur resulting from the resonant UV laser beam traversing the precession chamber to read out the \magHg{} spin precession.
Two different systematic effects, the vector and the direct light shifts, were considered. The vector light shift was measured for our previous experiment~\cite{Afach2014} using a \lampHg{} discharge lamp as the light source. This shift, which can be interpreted as the projection of the magnetic field of the photons traversing the precession chamber onto the \Bo-field direction, is magnetic-field direction dependent. As we exchanged the slightly off-resonant \lampHg{}~lamp with a laser beam resonantly locked to the \magHg{}~$6\,^1S_0 \rightarrow 6\,^3P_1~F=1/2$ transition, the shift reduced by a factor of 7.7~\cite{Fertl2013} to
\begin{equation}
	\begin{aligned}
		\Rr{}_{\rm VL}^{\uparrow} &=\left(1.5 \pm 6.9\right)\times10^{-7}~\text{and} \\ 
		\Rr{}_{\rm VL}^{\downarrow} &=\left(1.2 \pm 5.4\right)\times10^{-7},
		\label{eq:VecLightShift}
	\end{aligned}
\end{equation}
%{\color{blue}(I corrected the values you put, by multiplying by 3.84...)}
%%
where we kept the original uncertainties as the effect of the laser light power, which had an impact on the vector light shift, was not quantified.
In addition, the direct light shift accounts for the fact that while the probed atom is in the $6\,^3P_1~F=1/2$ state, the spin precesses at a different frequency. This will lead to a shift proportional to the light power and was estimated to be about 0.01 ppm~\cite{Abel2020} in the nEDM measurement with the same apparatus still using the \lampHg{}~discharge lamp. 
An increase in light power would also result in a decrease of the transverse depolarization time $T_2$ of the mercury precession, which was not observed.
Nevertheless, to account for a possible doubling of the light intensity because of the change to the resonant laser light, we estimate
\begin{equation}
		\Rr{}_{\rm DL}^{\uparrow/\downarrow} =\left(0.4\pm0.8\right)\times10^{-7}.
	\label{eq:DirLightShift}
\end{equation}
The contributions from both the vector and the direct light shifts are summed together and shown as the effect from the mercury light in Tab.~\ref{Tab_ErrorBudgetOnR}. 

Within the medium of spin-polarized \magHg{} vapor, UCNs are subjected to incoherent scattering on the \magHg{} nucleus, which can be described as a spin-dependent nuclear interaction. This acts as a pseudomagnetic field, which is proportional to the incoherent scattering length $\left|b_{\rm inc}\right| = \SI{15.5}{fm}$~\cite{Sears1992,Chanel2021}.
% of the neutrons on the \magHg{} atoms~\cite{Chanel2022}. 
For an imperfect $\pi/2$-pulse of the HgM, a residual polarization along \Bo{} creates a pseudomagnetic field, resulting in a frequency shift of UCNs and consequently a shift $\delta_{\rm inc}$ of \Rr{}. We estimated the random fluctuation of \magHg{} polarization and quantified the resultant error on \Rr{} as $\sigma_{R_{\rm inc}}^{\uparrow/\downarrow} \leq 5 \times 10^{-10}$. This effect is three orders of magnitude smaller than the light shift $\delta_{\rm light}$; hence, we consider it negligible.

The precession of spin-polarized particles is affected by magnetic-field fluctuations resulting from JNN, originating from thermal motion of charge carriers inside the electrodes. Because of the difference between adiabatic and nonadiabatic magnetic-field samplings for UCNs and \magHg~atoms, the volume-averaged fields sampled by both species are slightly different. A finite-element analysis was used to simulate temporal and spatial noise~\cite{Ayres2021b} from which we calculated the time-and-volume-averaged magnetic-field difference sensed by both particle ensembles. As JNN leads to random magnetic-field fluctuations independent of \Bo{} polarity, we assume that this effect $\delta_{\rm JNN}$ only increases the measurement uncertainty but does not shift the central \Rr{} value. The corresponding uncertainty was estimated to be $\sigma_{R_{\rm JNN}}^{\uparrow/\downarrow} \leq 1 \times 10^{-9}$. As this effect is two orders of magnitude smaller than $\delta_{\rm light}$, we did not include it in the error budget.

%Table \ref{Tab_ErrorBudgetOnR} summarizes the total error budget of \Rr{}. 
The analysis of the measured data, including all shifts and uncertainties as listed in Tab.\,\ref{Tab_ErrorBudgetOnR}, results in two independent \Rr{} values
\begin{equation} \label{Eq_FinalR}
	\begin{gathered}
			\begin{aligned}
					\Rr{}^{\uparrow} &= 3.8424563(08)_{\rm stat}(36)_{\rm sys} \enspace \text{and} \\
					\Rr{}^{\downarrow} &= 3.8424622(12)_{\rm stat}(59)_{\rm sys}
					%			\Rr{}^{\uparrow} = 3.8424528(66), \\
					%			\Rr{}^{\downarrow} = 3.8424621(68). 
				\end{aligned} 
		\end{gathered}
\end{equation}
at the limit of $\Gz{} = 0$. 
Both values are in agreement with our previous measurement of the neutron to mercury gyromagnetic ratio~\cite{Afach2014},
\begin{equation}
		\gN{}/\gHg{} = 3.8424574(30).
\label{eq:Ratio}
\end{equation}

\begin{table}[h!]
	\centering
	\begin{tabular}{|l|c|c|}
		\hline
		Effect / \num{1e-7} & $B_0$ up  & $B_0$ down \\
		\hline
		Statistics (uncertainty) & $\pm 8 $ & $\pm 12 $ \\ 
%		\hline
		Vertical magnetic-field gradient & $ 35 \pm 35 $ & $ -0.3 \pm 59 $ \\  
		Residual transverse field & $7.3 \pm 4.7 $ & $6.4 \pm 4.1 $ \\ 
		Mercury light & $1.9 \pm 6.9 $ & $1.6 \pm 5.5 $ \\  
%		\hline 
%		Total systematic effects & $ \pm 36 $ & $ \pm 59$ \\
		%\hline
		%\hline 
		%Negligible effects &&\\
		%Incoherent scattering & $\pm 4.73 \times 10^{-10}$ & $\pm 4.72 \times 10^{-10}$ \\
		%Johnson-Nyquist noise &$\pm 8.53 \times 10^{-10}$ & $\pm 8.51 \times 10^{-10}$ \\
		\hline	
	\end{tabular}
	\caption[]{Error budget for the overall errors on $\mathcal{R}$ resulting from statistics and from systematic effects for both \Bo{} directions. Note that the effects resulting from the vertical magnetic-field gradient and the transverse magnetic field were taken into account before the fit for the crossing point analysis.}
	\label{Tab_ErrorBudgetOnR}
\end{table}

%%%%%%%%%%%%%%%%%%%%%%%%%%%%%%%%%%%%%%%%%%%%%%%%%%%%%%%%%%%%%%%%%%%%%%%%%%%
\section{Interpretation of results} 

According to Eq.~\eqref{Eq_Rratio}, \bUcn{} was extracted at the limit of $\Gz{} = 0$ after correcting for all systematic effects $\delta_{\rm else}$, \begin{equation} \label{Eq_bUcnValue}
	\bUcn = \frac{\Rr{}^{\uparrow}-\Rr{}^{\downarrow}}{\Rr{}^{\uparrow}+\Rr{}^{\downarrow}}\left|\Bo\right| = \SI{-0.80}{\pico \tesla}, 
\end{equation} 
where $\left|\Bo\right| = 
%\left\langle\Bo{}\right\rangle = 
\SI{1037.19(2)}{\nano \tesla}$ was taken from the average \Bo{} value of all runs. The uncertainty of \bUcn{} was calculated by including uncertainties from $\Rr{}^{\uparrow}$, $\Rr{}^{\downarrow}$, and $\Bo{}$, 
%\begin{align} \label{Eq_bUcnError}
%	\sigma_{\bUcn} &= 
%		\left\lbrace \left[\frac{2 \Rr{}^{\downarrow} \left|\Bo\right|}{\left(\Rr{}^{\uparrow}+\Rr{}^{\downarrow}\right)^2} \sigma_{\Rr{}}^{\uparrow} \right]^2 
%		+ \left[\frac{-2 \Rr{}^{\uparrow} \left|\Bo\right|}{\left(\Rr{}^{\uparrow}+\Rr{}^{\downarrow}\right)^2} \sigma_{\Rr{}}^{\downarrow} \right]^2 \right. \nonumber \\
%		& \qquad \left.  + \left[ \frac{\Rr{}^{\uparrow}-\Rr{}^{\downarrow}}{\Rr{}^{\uparrow}+\Rr{}^{\downarrow}} \sigma_{\Bo{}} \right]^2 \right\rbrace ^{1/2}   
%		= \SI{1.73}{\pico \tesla}. 	
%\end{align}
\begin{equation} \label{Eq_bUcnError}
	\sigma_{\bUcn} = 
	\left\lbrace \left[\frac{2 \Rr{}^{\downarrow} \left|\Bo\right|}{\left(\Rr{}^{\uparrow}+\Rr{}^{\downarrow}\right)^2} \sigma_{\Rr{}}^{\uparrow} \right]^2 
	+ \left[\frac{-2 \Rr{}^{\uparrow} \left|\Bo\right|}{\left(\Rr{}^{\uparrow}+\Rr{}^{\downarrow}\right)^2} \sigma_{\Rr{}}^{\downarrow} \right]^2 
	+ \left[ \frac{\Rr{}^{\uparrow}-\Rr{}^{\downarrow}}{\Rr{}^{\uparrow}+\Rr{}^{\downarrow}} \sigma_{\Bo{}} \right]^2 \right\rbrace ^{1/2}   
	= \SI{0.96}{\pico \tesla}. 	
\end{equation}
The errors $\sigma_{\Rr{}}^{\uparrow}$ and $\sigma_{\Rr{}}^{\downarrow}$ were calculated by summing in quadrature the statistical error and all systematic effects that contributed to the error budget of \Rr{} (Tab.\,\ref{Tab_ErrorBudgetOnR}). $\sigma_{\Bo{}}$ was taken from the larger of the two standard deviations, $\sigma_{\Bo{}}^{\uparrow} = \SI{17}{pT} $ and $\sigma_{\Bo{}}^{\downarrow} = \SI{22}{pT}$, measured within one \Bo{} direction instead of taking the standard deviation of all 17 runs. 
Using Eq.~\eqref{Eq_bUcnDef}, $g_sg_p$ was derived as 
%\begin{widetext}
\begin{equation} \label{Eq_GsGp}
	g_sg_p = \bUcn \frac{2 \gN \mN H^2}{\hbar \lambda^2 \left[ H\left(\Nbot-\Ntop\right)-6\left\langle z \right\rangle\left(\Nbot+\Ntop\right) \right]}\left( 1-e^{\frac{-a}{\lambda}} \right)^{-1} \left( 1-e^{\frac{-H}{\lambda}} \right)^{-1}.	
\end{equation}
%\end{widetext}
With the measured \bUcn{} (Eq.~\eqref{Eq_bUcnValue}) and the estimated error $\sigma_{\bUcn}$ (Eq.~\eqref{Eq_bUcnError}), a 95\% confidence level limit on $g_sg_p$ gives
\begin{equation} \label{Eq_gsgpLim}
	g_sg_p\lambda^2 < 8.3 \times 10^{-28}\,\si{\meter\squared},
\end{equation} 
for $\SI{5}{\micro\meter} < \lambda < \SI{25}{\milli\meter}$. On the one hand, the upper limit of this range was defined as the thickness of the electrodes. Approaching the upper limit, the last two terms in Eq.~\eqref{Eq_GsGp} depart from 1, and the relation $g_sg_p \propto 1/\lambda^2$ is not fulfilled anymore, which reduces the measurement sensitivity on $g_sg_p$. On the other hand, the lower end of this range is constrained by the wavelength of UCNs and the surface property of the electrodes, such as the surface roughness, which was in the range of a few hundred nm, or the a-few-\si{um}-thick diamond-like-carbon coating that has a nucleon density between those of the aluminum and the copper electrodes.  

\begin{figure}[h!]
	\centering
	\includegraphics[width=.605\linewidth]{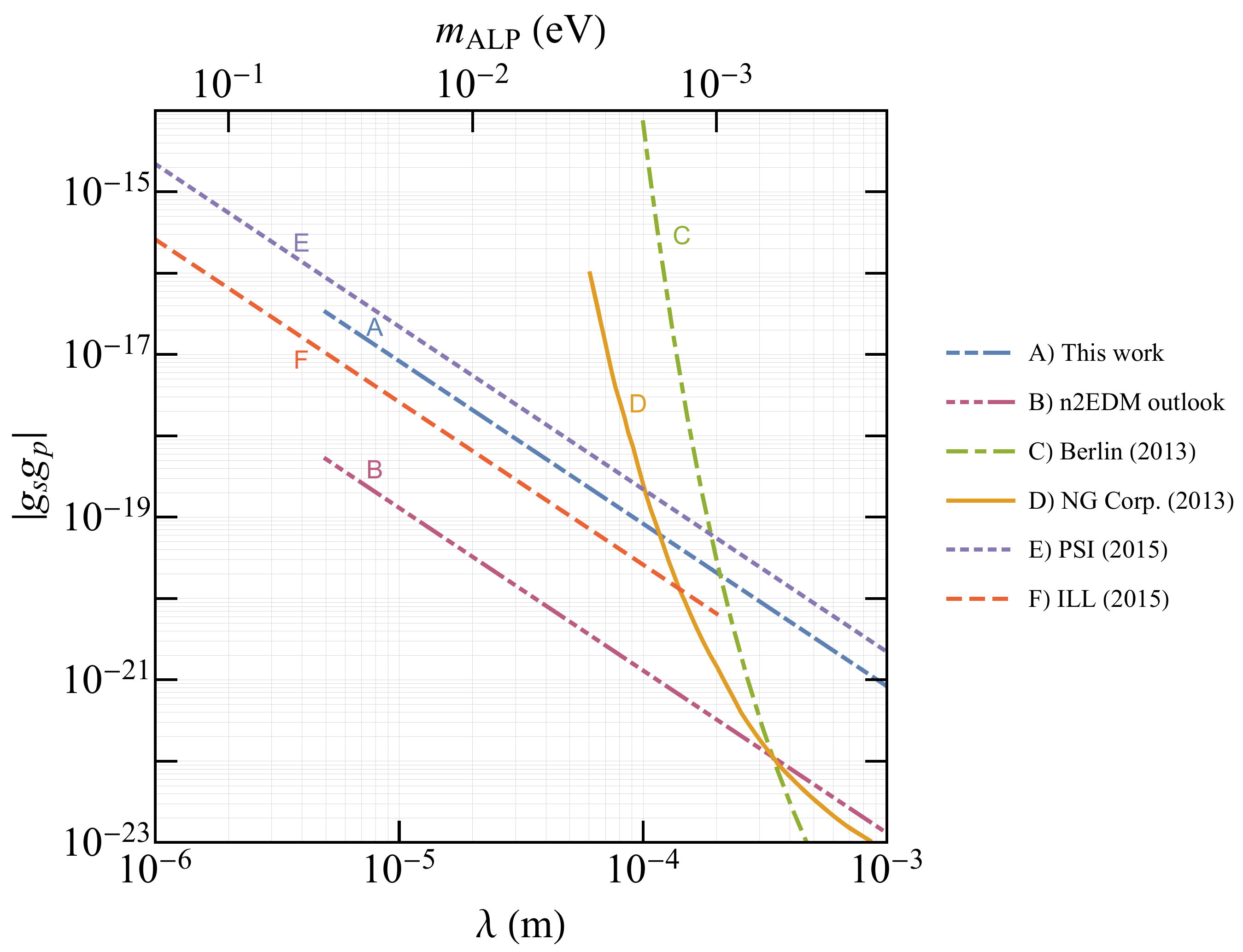}
	\caption[]{Upper limits on the $g_sg_p$-coupling of an interaction mediated by a spin-0 axion-like particle as a function of the interaction range $\lambda$ and the mass of an ALP \mAlps{}. A: this work (Eq.~\eqref{Eq_gsgpLim}). B: outlook on the n2EDM experiment with copper layers (Sec.\,\ref{Sec_AlpsN2EDM}). C: clock comparison of ${}^{3}{\rm He}$ and ${}^{129}{\rm Xe}$~\cite{Tullney2013}. D: clock comparison of ${}^{129}{\rm Xe}$ and ${}^{131}{\rm Xe}$~\cite{Bulatowicz2013}. E: our previous experiment with UCNs and \magHg{} atoms~\cite{Afach2015}. F: ${}^{3}{\rm He}$ depolarization~\cite{Guigue2015}.} 
	\label{Fig_ExclusionPlot}
\end{figure}

Figure~\ref{Fig_ExclusionPlot} shows the upper limits of $g_sg_p$ constrained by the most recent measurements covering an interaction range of $\SI{1}{\micro\meter} < \lambda < \SI{1}{\milli\meter}$. 
The upper horizontal axis displays the corresponding mass of an ALP \mAlps{}, with $\lambda = \hbar / \left(\mAlps c \right)$. 
The figure shows five measurement results, labeled from A to E\@. A is the limit obtained from this work (Eq.~\eqref{Eq_gsgpLim}), whereas E is obtained from our previous experiment~\cite{Afach2015}. 
Both experiments were based on a clock comparison of precession frequencies between polarized UCNs and \magHg{} atoms. An improvement on the sensitivity by a factor of 2.7 was accomplished, which is the current best limit of $g_sg_p$ obtained with UCNs. Additionally, we estimated the sensitivity of a new experiment, n2EDM, which is currently under construction at the PSI, to the SRSD interaction. The projected sensitivity is shown as B, and details of improvements on different statistical and systematic aspects are given in Sec.~\ref{Sec_AlpsN2EDM}\@. C is the result based on the comparison of spin-precession-frequency shifts of cohabiting ${}^{3}{\rm He}$ and ${}^{129}{\rm Xe}$ atoms, in the presence of an unpolarized mass of BGO (${\rm Bi}_{4}{\rm Ge}_{3}{\rm O}_{12}$) crystal~\cite{Tullney2013}. D results from the comparison of nuclear-magnetic-resonance-frequency shifts of cohabiting polarized ${}^{129}{\rm Xe}$ and ${}^{131}{\rm Xe}$ atoms in the presence of a nonmagnetic zirconium rod~\cite{Bulatowicz2013}. F is the result from the measurement of anomalous spin relaxation of polarized ${}^{3}{\rm He}$ atoms induced by an additional depolarization channel, which might be caused by the pseudomagnetic field generated from the $^3{\rm He}$-cell walls~\cite{Guigue2015}. 

\section{Prospects for ALP measurements in the \lowercase{n}2EDM experiment} \label{Sec_AlpsN2EDM}
A new experiment, n2EDM, to search for a permanent nEDM is currently under construction at the PSI\@. 
The aim is to search for an nEDM with a sensitivity of about \SI{1e-27}{\ecm}~\cite{Abel2019a,Ayres2021}. 
This will be accomplished by improved statistical sensitivity and an improved control of systematic effects. 
The apparatus can also be used to search for an SRSD interaction mediated by ALPs. 

The n2EDM apparatus features two precession chambers mounted on top of each other. 
This stack is made of three electrodes from aluminum spaced vertically by a height of \SI{12}{cm}.
In between the electrodes, cylindrical rings made of isolating polystyrene with a diameter of \SI{80}{cm}~\cite{Ayres2021} define the two precession chambers, increasing the volume of each precession chamber by a factor of three (a factor of six in total). 
The double-chamber design allows a simultaneous measurement in both chambers with opposite electric-field polarities while being exposed to the same magnetic-field direction.  
For ALP measurements where the electric field is not applied, observations simultaneously within both chamber are discussed. We estimate the improved statistical and systematic sensitivities with respect to this measurement. 

The statistical sensitivity of $\Rr{} = \fN{}/\fHg{}$ has contributions from neutron counting statistics and the uncertainty of the mercury-precession signal. With an optimized connection from the UCN source to the experiment and the double chamber with a factor of six larger total volume for UCNs, the projected number of detected neutrons after a free precession time of $\mathcal{T}=\SI{180}{s}$ will increase by a factor of eight. At the same time, the fringe visibility $\alpha$ slightly improves to $0.8$~\cite{Ayres2021}. Together, this will result in an improvement on the sensitivity of \fN{} by a factor of three, corresponding to a magnetic-field sensitivity of \SI{110}{fT}. The sensitivity of the mercury magnetometer is expected to be at least $\sigma_{B_{\rm Hg}} = \SI{30}{fT}$ per cycle~\cite{Ayres2021}. Hence, we expect a factor of six in statistical improvement to \SI{0.2}{ppm}. Recall that unexplained noise decreased the expected statistical sensitivity by a factor of 4--5 in the search presented in this article.

The largest uncertainty in the current result stems from the vertical magnetic-field gradient \Gz{}. Assuming that \Gz{} will be expanded up to fifth degree (Eq.~\eqref{Eq_Ggrav}), we expect the uncertainties of $G_{1,0}$, $G_{3,0}$, and $G_{5,0}$ determined by the HgM, the CsM array, and a more reproducible magnetic-field maps, respectively, to be~\cite{Ayres2021} 
\begin{equation}
	\begin{aligned}
		\sigma_{G_{1,0}} &\leq \sqrt{2}\, \sigma_{B_{\rm Hg}}/H' \approx \SI{2.4}{fT/cm}, \\
		\sigma_{G_{3,0}} &\leq  \SI{36e-3}{fT/cm^3},\enspace \text{and} \\
		\sigma_{G_{5,0}} &\leq  \SI{20e-6}{fT/cm^5},
	\end{aligned}
\end{equation}
where $H' = \SI{18}{cm}$ is the distance between the centers of the upper and the lower precession chambers.
%corresponding requirements on the accuracy and the sensitivity on $G_{1,0}$, $G_{3,0}$, and $G_{5,0}$ are discussed in Ref.~\cite{Ayres2021}. With the double-chamber design, we defined the {\it Top\textendash Bottom gradient} as 
%\begin{equation}
	%G_{\rm TB} = \frac{\left\langle B \right\rangle_{\rm Hg}^{\rm top}-\left\langle B \right\rangle_{\rm Hg}^{\rm bot}}{H'},
%\end{equation}  
%where $\left\langle B \right\rangle_{\rm Hg}^{\rm top}$ and $\left\langle B \right\rangle_{\rm Hg}^{\rm bot}$ are the average magnetic fields measured in the top and the bottom chambers by the HgM, respectively, and $H' = \SI{18}{cm}$ is the distance between the geometrical centers of the two chambers. $G_{\rm TB}$ will be accurately measured by the HgM\@. Assuming $G_{1,0} \sim G_{\rm TB}$, the uncertainty of $G_{1,0}$ could be estimated with the HgM sensitivity requirement $\sigma_{B_{\rm Hg}} = \SI{30}{fT}$ written as v. The accuracy requirements on the cubic ($G_{3,0}$) and the fifth-degree ($G_{5,0}$) gradients, which will be estimated with the CsM and the field maps, respectively, are elucidated in Sec.\,4.7 of Ref.~\cite{Ayres2021}. 
%The corresponding errors on these gradients was estimated accordingly.
Summing up all contributions, this implies a systematic uncertainty of $\sigma_{\Gz{}} = \SI{51}{fT/cm}$ on the vertical magnetic-field gradient; an improvement by a factor of 80 from the average uncertainty of \SI{4.05}{pT/cm} observed here.
%In ALP2015, we assumed $\Gz{} = G_{1,0}$, with $\sigma_{\Gz{}} \sim \SI{2.7}{pT/cm}$ (\Bup{}) and $\SI{3.0}{pT/cm}$ (\Bdo{})~\cite{Afach2014, Franke2013}. A factor of 53 (\Bup{}) and 59 (\Bdo{}) sensitivity gain can thus be expected for the systematic effect resulting from the vertical gradient $\sigma_{R_{\rm grav}}$, a factor  

The transverse magnetic field was estimated with field maps~\cite{Abel2022}.
%which gave errors of $\sigma_{\left\langle B_{\rm T}^2 \right\rangle} = \SI{0.5}{nT\tothe{2}}$ (\Bup{}) and $\SI{0.7}{nT\tothe{2}}$ (\Bdo{})~\cite{Afach2015}. 
In n2EDM, field maps will be used to measure all higher gradients above $G_{3,0}$, with a reproducibility requirement matching the previous repeatability~\cite{Ayres2021}.
This results in a tenfold improvement of the uncertainty $\sigma_{\left\langle B_{\rm T}^2 \right\rangle}$.
%We calculated the $\sigma_{\left\langle B_{\rm T}^2 \right\rangle}$ uncertainty reduction from ALP2015 to our current experiment (see Sec.\,\ref{subsec_crosspoint}), compared the reproducibility of $G_{5,0}$ measured from the current field maps and required for the n2EDM\@, and expect that the errors $\sigma_{R_{\rm T}}$ can be suppressed by a factor of 400 (\Bup{}) and 450 (\Bdo{}). 

By using a linearly polarized light scheme for reading out the mercury precession, we can suppress entirely the direct light shift. The vector light shift can be partially suppressed and characterized in dedicated measurements and will not significantly contribute to an overall error budget~\cite{CohenTannoudji1962}.
%A frequency shift of \fHg{}, depending on the intensity of the HgM probe light, will lead to an additional systematic effect on \Rr{}. In ALP2015, the \magHg{} atoms were pumped and probed by a $^{204}$Hg discharge lamp, whereas in the n2EDM, the HgM will be operated with a stabilized laser. With the requested stability and sensitivity for the HgM, a factor of 100 improvement on the mercury-light-induced systematic effect $\sigma_{R_{\rm light}}$ was assumed. 

In summary, this results in a measurement of \Rr{} with a statistical precision of about $\sigma_{\rm stat}^{\rm n2EDM}=\num{2e-7}$ and a systematic precision of about $\sigma_{\rm syst}^{\rm n2EDM}=\num{4e-8}$, a factor of 15 improvement compared to~\eqref{eq:Ratio}~\cite{Afach2014}.
With the estimated improvements on the statistical and, in particular, on the gradient-induced systematic uncertainties, a factor of three improvement to $g_sg_p\lambda^2 < 2.7 \times 10^{-28}\,\si{\meter\squared}$ is anticipated when using three electrodes all made of aluminum.
This might seem astonishing, considering the 25 times sensitivity gain on the measurement of the pseudomagnetic field estimated with Eq.~\eqref{Eq_bUcnError}. However, in the new experiment, the center-of-mass offsets estimating individually from each chamber are both $\langle z \rangle \approx \SI{0.1}{cm}$, which are a factor of four smaller, reducing the sensitivity by a factor of three.
%
%
%between the UCN and the mercury ensembles in the new apparatus. With the Monte Carlo simulations~\cite{Zsigmond2018, Bison2020}, the center-of-mass offsets were estimated to be $\left\langle z \right\rangle_{\rm top} = \SI{-0.09}{cm}$ and  $\left\langle z \right\rangle_{\rm top} = \SI{-0.12}{cm}$~\cite{Ayres2021}, which are much smaller than the value $\left\langle z \right\rangle = \SI{-0.235}{cm}$ obtained in ALP2015~\cite{Afach2014}. 

%The SRSD interaction, if exists, will only be observed between the nucleons on the electrodes and the UCN, whereas the effect on the \magHg{} atoms will be neglected. For an apparatus with both electrodes enclosing the relevant chamber made of the same material, the sensitivity of the influence from the pseudomagnetic field to the UCN ensemble decreases with a smaller $\left\langle z \right\rangle$ between both species.
Note that similar to the current experiment, using an asymmetric nucleon density between the upper and the lower boundary of each chamber, the sensitivity can be significantly increased. 
A possible approach might be placing copper sheets on the middle and the lower electrodes. 
%This will result in an approximately 44 times improvement for the detection of a spin-dependent short range force. 
%Copper, having a 3.3 times larger nucleon density compared to aluminum, placed at the bottom of each chamber will result in stronger effects on the UCN. 
With a \num{1}-\si{mm}-thick copper sheet, the interaction range up to \SI{1e-3}{m} can be covered. A new upper limit on the product of the couplings of about $g_sg_p\lambda^2 < 1.3 \times 10^{-29}\,\si{\meter\squared}$~(95\% C.L.) is then expected (marked as B in Fig.~\ref{Fig_ExclusionPlot}), when using copper having a 3.3 times larger nucleon density compared to aluminum.

\section{Conclusion}
This paper reports on the null result from a search for a hypothetical, short-range, spin-dependent interaction mediated by axion-like particles. Ultracold neutrons were stored simultaneously with \magHg{} atoms in a cylindrical chamber sandwiched between a copper and an aluminum electrode in a constant vertical magnetic field in the same apparatus used to measure the neutron electric dipole moment at the PSI\@. By measuring the precession-frequency ratio $\Rr{} = \fN{} / \fHg{}$ between UCNs and \magHg{} atoms in opposite magnetic-field directions, we searched for the SRSD interaction between nucleons of the electrodes and stored UCNs.

Systematic effects from magnetic-field gradients influenced the measurement of \Rr{}. The dominant effect arose from the center-of-mass offset between the two particle species in an effective vertical magnetic-field gradient \Gz{}. As \Rr{} is a linear function of \Gz{}, we intentionally applied a vertical magnetic-field gradient in the measurement and compared \Rr{} in different \Gz{} using the crossing-point analysis. 
%During the experiment, an array of \magCs{} magnetometer (CsM) were installed above and below the electrode stack to control and monitor the magnetic field. In addition, prior to this experiment, a set of magnetic-field maps of the apparatus were taken as part of our previous nEDM measurement. 
For a better estimation on \Gz{}, we combined both the online CsM data and magnetic-field maps taken at a different time using a dedicated device for mapping. The optimal method to incorporate both results was determined using synthesized data. 
%We concluded that by initially eliminating higher-degree-field components with degrees $l\geq 3$ and performing second-degree-polynomial fit to the remnant fields, a better control of the vertical-field gradient could be achieved with an 
%Using this method an accuracy of approximately 2 to \SI{3}{pT/cm} estimated from the toy models.
By applying this optimized method to measurement data, \Gz{} was estimated with an unprecedented precision of around \SI{4}{pT/cm}.  

By extracting \Rr{} at $\Gz{} = 0$ after correcting for all known systematic effects, a new limit on the product of the scalar and the pseudoscaler couplings, corresponding to the monopole-dipole interaction, gives $g_sg_p\lambda^2 < 8.3 \times 10^{-28}\,\si{\meter\squared}$ (95\% C.L.) in an interaction range of $\SI{5}{\micro\meter} < \lambda < \SI{25}{\milli\meter}$. 
This limit improves our previous experiment by a factor of 2.7, the best limit obtained with free neutrons. 

With the n2EDM apparatus at the PSI, we plan to search for a nonzero nEDM with a sensitivity of about \SI{1e-27}{\ecm}. By comparing the precession frequencies of \magHg{} atoms and the UCNs in the new spectrometer, a new, at least 15 times more accurate measurement of the gyromagnetic ratios $\gN{}/\gHg{}$ becomes possible. Further, a refined search of ALPs by placing a \num{1}-\si{mm}-thick copper layer on the corresponding bottom electrode of each chamber seems attractive. A new upper limit of $g_sg_p\lambda^2 < 1.3 \times 10^{-29}\,\si{\meter\squared}$ (95\% C.L.), a factor of 64 better than the result presented here, could be expected.

\section{Acknowledgments}
%The dataset was taken in 2014 - 2017 at PSI Villigen. 
We acknowledge the excellent support provided by the PSI
technical groups and by various services of the collaborating
universities and research laboratories. 
In particular, we acknowledge with gratitude the long term outstanding technical support by F.~Burri and M.~Meier. 
We thank the UCN source operation group BSQ for their support. We acknowledge financial support from the Swiss National
Science Foundation through projects 
No.\,117696,
No.\,137664,  
No.\,144473,
No.\,157079,
No.\,172626, 
No.\,126562,
No.\,169596,
No.\,178951~(all PSI),
No.\,181996~(Bern),
No.\,162574, 
No.\,172639~(both ETH).
%and No.\,140421~(Fribourg). 
%University of Bern acknowledges the support via the European Research Council under the ERC Grant Agreement No.\,715031-BEAM-EDM\@.  
%Contributions of the Sussex group have been made possible via STFC grants ST/M003426/1, ST/N504452/1, and ST/N000307/1. LPC Caen and
%LPSC Grenoble acknowledge the support of the French
%Agence Nationale de la Recherche (ANR) under Reference
%No.\,ANR-09-BLAN-0046 and the ERC Project No.\,716651-NEDM\@. 
The group from Jagiellonian University Cracow acknowledges the support from the National Science Center, Poland, through Grants No.\,UMO-2015/18/M/ST2/00056, No.\,UMO-2020/37/B/ST2/02349, and No.\,2018/30/M/ST2/00319, as well as by the Excellence Initiative – Research University Program at the Jagiellonian University.
%The Polish collaborators acknowledge support
%from the National Science Center, Poland, under grants
%No.\,UMO-2015/18/M/ST2/00056 and No.\,UMO-2020/37/B/ST2/02349
%P.M.\ acknowledges Grant No.\,SERI-FCS 2015.0594. 
%This work was also partly supported by
%the Fund for Scientific Research Flanders (FWO), and Project
%GOA/2010/10 of the KU Leuven.
This work was supported by the Research Foundation-Flanders (BE) under Grant No.\,G.0D04.21N.
We acknowledge the support
from the DFG (DE) on PTB core facility center of ultra-low magnetic field KO 5321/3-1 and TR 408/11-1.
We acknowledge funding provided by the Institute of Physics Belgrade through a grant by the Ministry of Education, Science and Technological Development of the Republic of Serbia.
This work is also supported by Sigma Xi grants \#\,G2017100190747806 and \#\,G2019100190747806, and by the award of the Swiss Government Excellence Scholarships (SERI-FCS) \#\,2015.0594. 
%In addition we are grateful
%for access granted to the computing grid PL-Grid infrastructure.

\bibliographystyle{NumEtAlNoTitleDOINew}
\bibliography{bib_ALPs}	

\end{document}